\documentclass[aps,prb,reprint,showpacs,twocolumn,nofootinbib,longbibliography]{revtex4-1}
\usepackage{amsmath,amssymb,bbm,hyperref,graphicx,color}
\usepackage[english]{babel}

\usepackage{txfonts}

\renewcommand{\Re}{\mathop{\mathrm{Re}}}
\renewcommand{\Im}{\mathop{\mathrm{Im}}}

\begin{document}

\title{Perturbative Field-Theoretical Renormalization Group Approach to
       Driven-Dissipative Bose-Einstein Criticality}

\author{Uwe C. T\"auber$^{1}$}
\author{Sebastian Diehl$^{2,3}$}

\affiliation{$^1$Department of Physics (MC 0435), Robeson Hall, 
  850 West Campus Drive, Virginia Tech, Blacksburg, VA 24061, USA}
\affiliation{$^2$Institute for Theoretical Physics, University of Innsbruck,
  A-6020 Innsbruck, Austria}
\affiliation{$^3$Institute for Quantum Optics and
  Quantum Information of the Austrian Academy of Sciences, A-6020 Innsbruck,
  Austria}

\date{\today} 

\begin{abstract} 
  The universal critical behavior of the driven-dissipative non-equilibrium 
  Bose-Einstein condensation transition is investigated employing the 
  field-theoretical renormalization group method. 
  Such criticality may be realized in broad ranges of driven open systems on 
  the interface of quantum optics and many-body physics, from 
  exciton-polariton condensates to cold atomic gases. 
  The starting point is a noisy and dissipative Gross-Pitaevski equation 
  corresponding to a complex valued Landau-Ginzburg functional, which captures
  the near critical non-equilibrium dynamics, and generalizes Model A for 
  classical relaxational dynamics with non-conserved order parameter.  
  We confirm and further develop the physical picture previously established 
  by means of a functional renormalization group study of this system.
  Complementing this earlier numerical analysis, we analytically compute the
  static and dynamical critical exponents at the condensation transition to
  lowest non-trivial order in the dimensional $\epsilon$ expansion about the
  upper critical dimension $d_c = 4$, and establish the emergence of a novel
  universal scaling exponent associated with the non-equilibrium drive.
  We also discuss the corresponding situation for a conserved order parameter 
  field, i.e., (sub-)diffusive Model B with complex coefficients. 
 \end{abstract}

\maketitle

\section{Introduction} 
\label{sec:intro}

Experimental systems that are characterized by a strong coupling of light to a 
large number of matter degrees of freedom \cite{carusotto13} hold the 
potential of developing into laboratories for non-equilibrium statistical 
mechanics, where phase transitions among stationary states far away from 
thermodynamic equilibrium could be studied.  
Instances of such systems have recently been demonstrated in a variety of 
contexts:  
In ensembles of ultracold atoms, Bose-Einstein condensates (BEC) placed in 
optical cavities have allowed to achieve strong light-matter coupling, and led
to the realization of open Dicke models 
\cite{esslingerdicke,esslingerQEDreview}.
The corresponding phase transition has been studied in real time, including 
the determination of the associated critical exponent \cite{esslingerdicke3}.  
In systems of trapped ions \cite{blatt12}, Ising models with variable-range 
interactions of a few hundred quantum spins have been created 
\cite{britton12}. 
Other platforms, which hold the promise of being developed into true many-body 
systems by scaling up the number of presently existing elementary building 
blocks in the near future, are provided by arrays of microcavities 
\cite{clarke-nature-453-1031,hartmann08,houck12,koch13}, and also 
optomechanical setups \cite{marquardt09,chang11,ludwig13}.

Genuine many-body ensembles in the above class are furthermore realized in the 
context of pumped semiconductor quantum wells in optical cavities 
\cite{imamoglu96}. 
Here, non-equilibrium Bose-Einstein condensation of exciton-polaritons has 
been achieved \cite{Kasp2006,lagoudakis08,Roumpos24042012} -- the effective 
bosonic degrees of freedom result from a strong hybridization of cavity light 
and excitonic matter states \cite{SnokeBook,keeling10,carusotto13}.  

All these systems exhibit the crucial ingredients for non-trivial critical 
scaling behavior at a continuous non-equilibrium phase transition. 
This triggers broader theoretical questions on the actual nature and possible
universality classes of such non-equilibrium critical points. 
At first sight, invoking the concept of universality, implying a huge ``loss 
of memory'' on details of the microscopic physics, it may seem questionable 
whether the microscopic non-equilibrium conditions will result in any 
physically observable consequences at the macroscopic level at all. 
In particular, for equilibrium dynamical critical behavior there exists a 
well-developed theoretical framework based on the seminal work of Hohenberg 
and Halperin (HH) \cite{hohenberg77:_theor} (and other authors), who 
classified various types of dynamical critical behavior into diverse 
equilibrium dynamical universality classes, known as Models A to J, depending 
on the conserved or non-conserved nature of the order parameter itself, and on
its dynamical couplings to other slow conserved modes. 

However, there are two key ingredients, shared by the systems described above, 
that place these many-body ensembles apart from equilibrium systems and may in
fact cause novel universal physical features. 
First, they are strongly driven by external fields, such as coherent 
electromagnetic radiation provided by lasers, and undergo a cascade of internal 
relaxation mechanisms \cite{carusotto13}. 
This  non-equilibrium drive and balancing dissipation adds to the 
Hamiltonian dynamics, and causes both reversible (coherent) 
and irreversible (dissipative) dynamics to appear on an equal footing, albeit 
originating from physically distinct and independent mechanisms. 
In turn, this induces manifest violations of the detailed-balance conditions 
characteristic of a many-body system in thermal equilibrium. 
Indeed, these drive-induced non-equilibrium perturbations transcend mere 
violations of the Einstein relations (or fluctuation-dissipation theorem) that
in equilibrium connect relaxation coefficients with associated thermal noise
strengths:
Such perturbations have been found to generically become irrelevant in the
vicinity of a second-order phase transition (for a concise overview, see 
Ref.~\onlinecite{tauber02}).
Second, these systems are characterized by the absence of the conservation of 
particle number. 
This is due to the admixture of light to the matter constituents, which 
opens up strong loss channels for the effective hybrid light-matter degrees of 
freedom, in turn making it necessary to counterpoise these losses by 
continuous pumping in order to achieve stable stationary states.

Deciding whether or not these ingredients indeed cause universal behavior 
distinct from equilibrium motivates -- and in fact necessitates -- a thorough 
theoretical analysis of the nature of criticality in such non-equilibrium 
quantum systems. 
A key representative of potential non-equilibrium criticality is provided by 
the driven-dissipative Bose-Einstein condensation transition, relevant to the 
experiments with exciton-polariton condensates described above. 
For such systems, indeed, a new independent critical exponent associated with 
the non-equilibrium drive has recently been identified within a functional 
renormalization group (RG) approach \cite{sieberer13:_dynam,sieberer13_long}. 
This exponent describes universal decoherence at long distances, and is 
observable, e.g., in the momentum- and frequency-resolved single-particle 
response, as probed in homodyne detections of exciton-polariton systems 
\cite{utsu08}. 
Furthermore, an effective thermalization mechanism for the low-frequency 
distribution function has been found, reflected in an emergent symmetry at the 
(classical, equilibrium) Wilson-Fisher fixed point. 

In this work, we employ the field-theoretical RG 
\cite{amitbook,itzyksonbook,zinnjustinbook,kleinert-book} in a perturbative 
dimensional $\epsilon$ expansion for a complementary study of driven 
Bose-Einstein criticality. 
Here $\epsilon = 4 - d$ measures the distance from the upper critical 
dimension $d_c = 4$, and serves as the effective small parameter in the 
perturbation series. 
In this framework, we confirm yet also further develop the physical picture 
obtained previously within the functional RG approach. 
Both our perturbative two-loop and the non-perturbative functional RG analysis
\cite{sieberer13:_dynam,sieberer13_long} are based on an effective 
long-wavelength description in terms of a noisy Gross-Pitaevskii equation with 
complex coefficients 
\cite{keeling04,carusotto05,szymanska06,wouters07,keeling08}, which in turn
constitutes a variant of the time-dependent complex Ginzburg-Landau equation 
for a two-component order parameter field \cite{DeDominicis75}.
Such complex stochastic differential equations have also found extensive 
applications in the modeling of spontaneous structure formation in
non-equilibrium systems \cite{cross93,crossbook}. 
Remarkably, these comprise coupled non-linear oscillators subject to external
noise near a Hopf bifurcation instability \cite{risler05}, and even spatially 
extended evolutionary game theory and the dynamics of cyclically competing 
populations \cite{frey10}.

We present a concise account of the main results of this work in the 
following Sec.~\ref{sec:results}.
The  remainder of this paper is organized as follows: 
In Sec.~\ref{sec:model}, we explain the microscopic model based on a 
stochastic Gross-Pitaevskii equation with complex coefficients, and introduce 
the equivalent dynamical response functional integral as appropriate for a 
subsequent diagrammatic evaluation.
We also discuss the relationship of our model with Model E that governs the 
equilibrium critical dynamics of planar ferromagnets and the normal- to 
superfluid phase transition.
Section~\ref{sec:renor} comprises the bulk of this work.
It contains an explanation of the renormalization scheme employed to deal with
the emergent ultraviolet  (UV) divergences, and details how the 
critical scaling properties in the infrared (IR) region may 
subsequently be obtained from solutions of the associated RG flow equations.
Finally, Sec.~\ref{sec:concl} offers a summary and concluding remarks, and
supplementary appendices provide more technical details.

\section{Key Results and Physical Picture}
\label{sec:results}

\emph{Generalized dynamic scaling forms.} --
Near the continuous condensation transition for driven-dissipative boson 
system, we {\em derive} generalized scaling laws for the dynamic response and
correlation functions at wavevector $\mathbf{q}$ and frequency $\omega$:
\begin{eqnarray} \label{eq:chs}
\chi(\mathbf{q},\omega,\tau)^{-1} &\propto& |\mathbf{q}|^{2 - \eta} 
\left( 1 + i a \, |\mathbf{q}|^{\eta - \eta_c} \right) \\
&& \times \, {\hat \chi}\left( \frac{\omega}{|\mathbf{q}|^z 
\left( 1 + i a \, |\mathbf{q}|^{\eta - \eta_c} \right)} , 
|\mathbf{q}| \, \xi \right)^{-1} \ , \nonumber \\
C(\mathbf{q},\omega,\tau) &\propto& |\mathbf{q}|^{- 2 - z + \eta'} 
{\hat C}\left( \frac{\omega}{|\mathbf{q}|^z} ,  |\mathbf{q}| \, \xi , 
a |\mathbf{q}|^{\eta - \eta_c} \right) \ , \label{eq:cos}
\end{eqnarray}
where $a$ is a non-universal constant, and the correlation length diverges as 
$\xi \propto |\tau|^{- \nu}$ as the critical point is approached,
$\tau \propto T - T_c \to 0$.
Here, $\nu$, $\eta$, and $z$ represent the standard {\em equilibrium} static 
and dynamical critical exponents, while $\eta_c$ and $\eta'$ constitute novel 
scaling exponents induced by the non-equilibrium drive and associated 
potential violation of detailed balance. 

The origin of these new scaling exponents is immediately transparent from 
the description of the problem in terms of a Janssen-De~Dominicis (or 
Martin-Siggia-Rose) functional integral \cite{janssen76,bausch76,dominicis76}: 
Owing to the competition of coherent and dissipative dynamics in the driven 
problem, {\em two} independent mass scales appear, as compared to a single one
in the closely related, purely relaxational Model A  in 
equilibrium critical dynamics.
This causes a more complex critical scenario akin to a bicritical point.
In addition, the fluctuation-dissipation theorem that relates the dynamic
response with the correlation function in thermal equilibrium (for which
$\eta' = \eta$ are hence identical) is in general violated.
Indeed, there arise new ultraviolet divergences, specifically for two 
couplings that are marginal at the Gaussian fixed point.

\emph{Asymptotic thermalization.} --
We establish that the renormalization group flow, already to one-loop order, 
drives the system towards an effectively {\em equilibrium} fixed point, where
detailed balance is satisfied.
The fluctuation-dissipation theorem then implies that
\begin{equation} \label{eq:eta'}
\eta' = \eta
\end{equation}
holds {\em exactly} for the Fisher exponents that characterize the anomalous 
algebraic spatial decay of the order parameter dynamic response 
\eqref{eq:chs} and correlation \eqref{eq:cos} functions at criticality.

The analysis of the two-loop RG flow equations furthermore yields that the
asymptotic fixed point values of all non-equilibrium coupling parameters
induced by the external drive vanish.
Consequently the static critical exponents are precisely those of the 
equilibrium $O(2)$-symmetric Ginzburg-Landau-Wilson Hamiltonian (XY model), 
namely
\begin{eqnarray} \label{eq:enu}
&&\nu = \frac12 + \frac{\epsilon}{10} + O(\epsilon^2) \ , \\
&&\eta = \frac{\epsilon^2}{50} + O(\epsilon^3) \label{eq:eta}
\end{eqnarray}
to lowest non-trivial order in the dimensional $\epsilon$ expansion, and the
associated dynamic critical exponent is the standard one for the 
$O(2)$-symmetric Model A that describes the purely relaxational kinetics of a 
non-conserved two-component order parameter field \cite{bausch76},
\begin{equation} \label{eq:exz}
z = 2 + c \eta \ , \quad c = 6 \ln \tfrac43 - 1 + O(\epsilon) \ .
\end{equation}
Thus, the system's asymptotic long-wavelength and low-frequency properties 
become effectively {\em thermalized}; see also Ref.~\onlinecite{risler05}. 
Yet a novel universal scaling exponent appears in the subleading scaling 
behavior, see Eqs.~\eqref{eq:chs} and \eqref{eq:cos}, which originates from 
the driven-dissipative setup \cite{DeDominicis75}.

\emph{Novel drive exponent.} -- 
Despite the fact that the system approaches an effectively equilibrium RG 
fixed point, one of the marginal couplings induces a novel and independent 
(but subleading) scaling exponent that captures the fadeout of coherent 
quantum fluctuations relative to their thermal, dissipative counterparts.
We compute the new drive exponent to second order in the loop expansion, i.e., 
to order $\epsilon^2$: 
\begin{equation} \label{eq:etac}
\eta_c = c' \eta \ , \quad c' = - \left( 4 \ln \tfrac43 - 1 \right) 
+ O(\epsilon) \ ,
\end{equation}
Eq.~\eqref{eq:etac} is one of the central results of this paper. 
Together with the asymptotic thermalization, these key findings corroborate 
the earlier functional RG study of 
Ref.~\onlinecite{sieberer13:_dynam,sieberer13_long}.
They also underscore the well-known remarkable stability of Model A with 
respect to non-equilibrium perturbations \cite{haake84,grinstein85,bassler94}.

\emph{Hierarchical structure of non-equilibrium criticality.} --
In this way, we confirm the hierarchical structure of the model's critical 
behavior, in the following sense: 
The static critical behavior is characterized by the $O(2)$ universality 
class, described by the rotationally invariant Ginzburg-Landau-Wilson 
Hamiltonian for a two-component order parameter field.
In equilibrium dynamical criticality, the static properties are supplemented 
-- but not modified -- by the dynamical critical exponent, e.g. 
Eq.~\eqref{eq:exz} for Model A.
Yet here, the non-equilibrium conditions give rise to the new and independent 
scaling exponent \eqref{eq:etac}. 
We thus establish the following pattern:  
While the non-equilibrium drive modifies neither the universal $O(2)$ static 
nor even the dynamical critical behavior of Model A, it still adds novel 
universal scaling features.
This situation is reminiscent of (but different from) the emergence of a new
critical exponent in Model A associated with the non-equilibrium relaxation 
following a sudden temperature quench from random initial conditions to the
critical point \cite{janssen89,calabrese05}.

\emph{Relation to equilibrium dynamic criticality.} -- 
We elaborate on the relation of the driven Bose-Einstein condensation to its 
equilibrium counterpart, which is described by Model E in the terminology of 
HH.
The hydrodynamic conservation laws relevant for the latter model have two 
crucial consequences, which set it apart from our non-equilibrium situation: 
First, the dynamical critical exponent is modified due to the existence of a 
new relevant reversible coupling to a diffusive mode, and fixed by rotational 
invariance in order parameter space to $z = d / 2$ (in the strong dynamic 
scaling regime), quite distinct therefore from our result for the relaxational 
dynamical critical exponent coinciding with Model A. 
Second, these conservation laws exclude the addition of a second mass scale 
to the problem, and in consequence, there emerges no counterpart of the drive 
exponent in equilibrium Bose-Einstein condensation criticality. 

Intriguingly, the additional drive exponent is absent for a Model B version
(in the HH classification) of the complex Ginzburg-Landau equation.
Instead of the purely relaxational kinetics of a non-conserved order 
parameter, in this situation a diffusive relaxation for a conserved order 
parameter field is implemented. 
Thermalization along with the conservation law and ensuing structure of the 
non-linear relaxation vertices now imply the {\em exact} scaling laws 
\eqref{eq:eta'} and \cite{bausch76}
\begin{equation} \label{eq:exzb}
z = 4 - \eta \ . 
\end{equation}
In addition, we establish the identity
\begin{equation} \label{eq:etacb} 
\eta_c = \eta + O(\epsilon^3) \ ,
\end{equation} 
at least to two-loop order.

\emph{Theoretical approach and renormalization scheme.} --  
The perturbative field-theoretical RG approach 
\cite{amitbook,itzyksonbook,zinnjustinbook,kleinert-book} provides a 
well-established tool for the quantitative characterization of critical 
behavior close to the upper critical dimension $d_c = 4$; more precisely, it 
is perturbatively controlled in the dimensional parameter $\epsilon = 4 - d$.
In equilibrium it has moreover been demonstrated that the structure of the RG
flow equations and the ensuing universality classes remain robust even in 
extensions down to three dimensions.
In this paper we work at lowest non-trivial order in $\epsilon$, i.e., to 
$O(\epsilon)$ for, e.g., the correlation length exponent $\nu$ and the 
fixed-point value of the non-linear coupling $u$, but to order $\epsilon^2$ 
for the Fisher, dynamic, and drive exponents, whose anomalous scaling 
dimensions require a calculation at the two-loop level. 
A key advantage of this approach in the context of non-equilibrium criticality 
is the possibility of a direct and quantitative comparison of our findings 
with the well-known results for equilibrium dynamical criticality displayed by 
the phenomenological models of HH \cite{hohenberg77:_theor}, which have not 
yet been comprehensively studied in a functional RG framework (except for 
Refs.~\onlinecite{canet07,canet11:_gener,mesterhazy13}). 

We remark that the field-theoretical RG approach differs conceptually from the 
functional RG based on Wetterich's equation \cite{wetterich93:_exact}: 
The latter constitutes an exact reformulation of a given functional integral 
in terms of a functional differential equation, in this way at least in 
principle addressing the full many-body problem 
\cite{berges02:_nonper,pawlowski07,rosten12,boettcher12,RevModPhys.84.299}. 
Critical behavior can then be studied by a suitable fine-tuning of parameters. 
In contrast, the field-theoretical RG focuses immediately on the critical 
surface of the problem, in this way isolating the universal critical behavior 
from the outset (see, e.g., 
Refs.~\onlinecite{amitbook,itzyksonbook,zinnjustinbook,kleinert-book}; and for
the application to dynamic critical phenomena Refs.~\onlinecite{DeDominicis75,bausch76,vasilievbook,folk06,kamenevbook,tauberbook}). 
While, therefore, non-universal aspects of the problem are projected out, it 
provides a perhaps more fundamental understanding of the emergence of scaling 
properties, and moreover allows us to obtain explicit analytical results for 
the critical exponents, cf. Eqs.~\eqref{eq:enu}-\eqref{eq:etac}. 

In short, by means of the field-theoretical RG approach we provide 
complementary strong evidence that the microscopic non-equilibrium character 
bears observable consequences up to the largest  distance and 
time scales in driven-dissipative Bose-Einstein condensation.

\section{The Model} 
\label{sec:model}

\subsection{The dissipative Gross-Pitaevskii equation with noise}

Driven-dissipative Bose-Einstein condensation in exciton-polariton systems is 
properly described by a noisy dissipative Gross-Pitaevskii equation with 
complex coefficients 
\cite{keeling04,carusotto05,szymanska06,wouters07,keeling08},
\begin{eqnarray} \label{eq:gpe}
&&i \partial_t \psi (\mathbf{x},t)= \Bigl[ - \left( A - i D \right) \nabla^2 
- \mu + i \chi \nonumber \\
&&\qquad\qquad\quad + \left( \lambda - i \kappa \right) |\psi(\mathbf{x},t)|^2 
\Bigr] \, \psi(\mathbf{x},t) + \zeta(\mathbf{x},t) \ . 
\end{eqnarray}
It basically coincides with the time-dependent complex Ginzburg-Landau 
equation, which has been prominently employed to describe pattern formation in 
non-equilibrium systems, typically however in the deterministic 
limit without noise \cite{cross93,crossbook}.
A stochastic variant has been analyzed in the context of coupled anharmonic
oscillators \cite{risler05}.
Here, the complex bosonic field $\psi$ describes the polariton degrees of 
freedom. 
The complex coefficients have clear physical meanings; in particular, 
$\chi = (\gamma_p - \gamma_l)/2$ is the net gain, i.e., the balance of the 
incoherent pump rate $\gamma_p$ and the local single-particle loss rate 
$\gamma_l$.
The positive parameters $\kappa$ and $\lambda$ represent the two-body loss and 
interaction strength, respectively, while $A = 1/2m_\text{eff}$ relates to the 
effective mass of the polaritons. 
Typically, this equation is not presented with an explicit diffusion 
coefficient $D$, but rather with a frequency dependent pump term 
$\sim \eta \partial_t \psi$ adding to the left hand side of the equation 
\cite{wouters_freq10,wouters_freq210}. 
The form \eqref{eq:gpe} is then recovered upon division by $1 - i \eta$, i.e., 
with $D = A \eta$ and subleading corrections to the other coefficients which 
are complex to begin with. 
We emphasize that, due to the freedom of normalizing the time derivative term  
as above in the equation of motion, this model accurately captures the physics 
close to the phase transition, since it describes the most general 
low-frequency dynamics in a systematic derivative expansion that incorporates 
all relevant couplings (in the sense of the RG) in dimensions $d > 2$.
Finally, the noise described by the fluctuating complex variable $\zeta$ is 
taken to be Gaussian, white, and Markovian,  and hence is fully 
characterized by the correlators 
\begin{eqnarray} \label{eq:noi}
&&\left\langle \zeta^*(\mathbf{x},t) \right\rangle 
= \left\langle \zeta(\mathbf{x},t) \right\rangle = 0 \ , \nonumber \\
&&\left\langle \zeta^*(\mathbf{x},t) \, \zeta(\mathbf{x}',t') \right\rangle 
= \gamma \, \delta(\mathbf{x} - \mathbf{x}') \, \delta(t - t') \ , \\ 
&&\left\langle \zeta^*(\mathbf{x},t) \, \zeta^*(\mathbf{x}',t') \right\rangle 
= \left\langle \zeta(\mathbf{x},t) \, \zeta(\mathbf{x}',t') \right\rangle  = 0 
\ . \nonumber
\end{eqnarray}

Physical stability requires $A,D,\lambda$, and $\kappa$ to be positive. 
The parameter $\chi$ is negative in the disordered phase, where the global 
$U(1)$ gauge symmetry (or $O(2)$ rotational symmetry in the complex plane) is 
not spontaneously broken. 
Our calculations will be carried out in this regime, i.e., we approach the 
phase transition from the disordered side, in contrast to the non-perturbative
RG analysis in Refs.~\onlinecite{sieberer13:_dynam,sieberer13_long}. 
For increasing pump rate $ \gamma_p$, the gain $\chi$ eventually turns 
positive and the system undergoes a continuous, driven Bose condensation 
transition: 
The instability occuring for a state with vanishing polariton field 
expectation value is cured by the expression of a polariton condensate 
$\langle \psi(\mathbf{x},t) \rangle \neq 0$. 
The parameter $\mu$, which effectively assumes the role of a chemical 
potential, is fixed by the requirement of stationarity, see below. 
The Langevin equation \eqref{eq:gpe} can be formally derived from a 
microscopic description in terms of a quantum master equation (see, e.g., 
Ref.~\onlinecite{zollerbook}) upon employing canonical power counting in the 
vicinity of the critical point \cite{sieberer13:_dynam,sieberer13_long}. 

For the analysis of the critical behavior, it turns out useful to introduce the 
following ratios (and sign conventions):
\begin{equation} \label{eq:rat}
r = - \frac{\chi}{D} \, , \;\ r'= - \frac{\mu}{D} \, , \;\ 
u' = \frac{6 \kappa}{D} \, , \;\ r_K = \frac{A}{D} \, , \;\ 
r_U  = \frac{\lambda}{\kappa} \ .
\end{equation}
The relaxation rate $D$ may be small on the microscopic scales of actual
experiments, but it plays a key role for the dominantly diffusive dynamics in 
the vicinity of the phase transition, as will be confirmed in the subsequent
calculation. 
Therefore, it is useful to express all external control parameters relative to 
$D$. 
The quantities $r_K$ and $r_U$ are dimensionless, giving the ratio of real and 
imaginary parts of the couplings in Eq.~\eqref{eq:gpe}, and therefore 
describing the relative strength of coherent vs. dissipative dynamics. 
In these units Eq.~\eqref{eq:gpe} takes the form
\begin{eqnarray} \label{eq:gpe2}
\partial_t \psi(\mathbf{x},t) &=& - D \, \Bigl[ r + i r' 
- \left( 1 + i r_K \right) \nabla^2 \nonumber \\
&&+ \frac{u'}{6} \left( 1 + i r_U \right) |\psi(\mathbf{x},t)|^2 \Bigr] \, 
\psi(\mathbf{x},t) + \xi(\mathbf{x},t) \quad \\
&=& - D \, \frac{\delta \bar H[\psi]}{\delta \psi^*(\mathbf{x},t)} 
+ \xi(\mathbf{x},t) \ , \label{eq:gpe3}
\end{eqnarray}
with the stochastic noise $\xi(\mathbf{x},t) = - i \zeta(\mathbf{x},t)$ 
governed by the same correlations \eqref{eq:noi} as $\zeta$. 
Formally, as indicated in Eq.~\eqref{eq:gpe3}, this describes the relaxational 
kinetics of Model A with a non-conserved order parameter, however with a 
non-Hermitean effective ``Hamiltonian''
\begin{eqnarray} \label{eq:ham}
&&\bar H[\psi] = \int d^dx \, \bigg[ \left( r + i r' \right) 
|\psi(\mathbf{x},t)|^2 + \left( 1 + i r_K \right) 
|\nabla \psi(\mathbf{x},t)|^2 \nonumber \\
&&\qquad\qquad\qquad\quad + \frac{u'}{12} \left( 1 + i r_U \right) 
|\psi(\mathbf{x},t)|^4 \biggr] \ .
\end{eqnarray}
The complex coefficients in Eq.~\eqref{eq:ham} reflect the presence of the 
non-equilibrium drive.
The theory becomes critical (massless) when $r, r' \to 0$ (more precisely, the
renormalized counterparts of these parameters $\tau, \tau' \to 0$) 
simultaneously.
We remark that Ref.~\onlinecite{risler05} addressed the distinct physical 
situation where the uncoupled oscillation frequence $r'$ was held fixed.
The calculation was then performed in a rotating reference frame, which
formally amounts to setting $r' = 0$ in our analysis.

\subsection{Field theory representation}

\begin{figure*}[h,t]
\includegraphics[width=1.8\columnwidth]{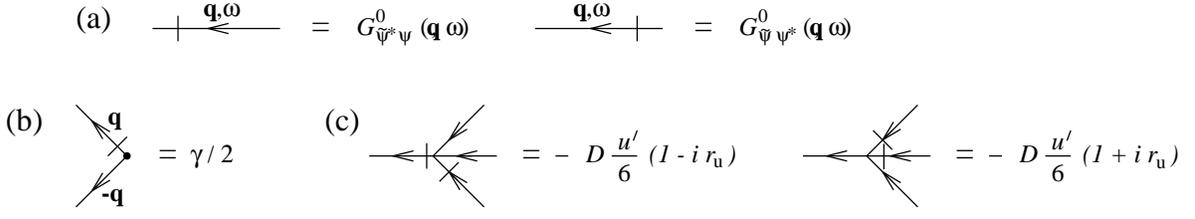}
\caption{Elements of the diagrammatic perturbation expansion: 
(a) Bare retarded and advanced propagators \eqref{eq:prr}, where the arrows
    reflect the causal temporal flow from $\psi$ to $\tilde\psi$ fields, while
    the perpendicular bars indicate complex conjugation; 
(b) two-point noise vertex; and (c) four-point relaxation vertices (note that
    vertex functions are depicted with truncated external legs).}
\label{fg:fig1} 
\end{figure*}
The nonlinear partial differential equation \eqref{eq:gpe} or \eqref{eq:gpe3} 
represents a classical stochastic evolution, which is readily mapped into an 
equivalent Janssen-De~Dominicis (or Martin-Siggia-Rose) functional integral 
representation \cite{janssen76,bausch76,dominicis76}; for detailed explanations,
see, e.g., Refs.~\onlinecite{tauber07,vasilievbook,kamenevbook,tauberbook}.
This formulation renders it amenable to straightforward perturbative expansions
with respect to the nonlinear coupling $u'$, the use of diagrammatic 
techniques, and subsequent implementation of the field-theoretical dynamical RG.
The Janssen-De~Dominicis response functional corresponding to 
Eq.~\eqref{eq:gpe3} with noise correlations \eqref{eq:noi} reads 
\begin{eqnarray} \label{eq:act}
&&\! \mathcal{A}[\tilde\psi,\psi] = \int \! dt \, d^dx \, \biggl[ 
\tilde\psi^*(\mathbf{x},t) \, \biggl( \partial_t \psi(\mathbf{x},t) + D \, 
\frac{\delta \bar H[\psi]}{\delta \psi^*(\mathbf{x},t)} \biggr) \nonumber \\
&&\qquad\qquad\qquad\qquad\quad + \ \text{h.c.} \ - \frac{\gamma}{2} \, 
|\tilde\psi^*(\mathbf{x},t)|^2 \biggr] \\
&&= \int \!\! dt \, d^dx \, \biggl[ \tilde\psi^*(\mathbf{x},t) \left( 
\partial_t + D \left[ r + i r' - \left( 1 + i r_K \right) \nabla^2 \right] 
\right) \psi(\mathbf{x},t) \nonumber \\
&&\qquad\qquad\;\ + \tilde\psi(\mathbf{x},t) \left( \partial_t + D \left[ r 
- i r' - \left( 1 - i r_K \right) \nabla^2 \right] \right) 
\psi^*(\mathbf{x},t) \nonumber \\
&&\qquad\quad + D \, \frac{u'}{6} \left( 1 + i r_U \right) 
\tilde\psi^*(\mathbf{x},t) \, |\psi(\mathbf{x},t)|^2 \, \psi(\mathbf{x},t) 
\nonumber \\
&&\quad\ + D \, \frac{u'}{6} \left( 1 - i r_U \right) \tilde\psi(\mathbf{x},t)
\, |\psi(\mathbf{x},t)|^2 \, \psi^*(\mathbf{x},t) - \frac{\gamma}{2} \, 
|\tilde\psi^*(\mathbf{x},t)|^2 \biggr] \ . \nonumber
\end{eqnarray}
It provides the statistical weight $P[\psi] \propto 
\int \mathcal{D}[i \tilde{\psi}] \, e^{- \mathcal{A}[\tilde\psi,\psi]}$ for
the stochastic process encoded in the Langevin equation \eqref{eq:gpe2} for 
$\psi(\mathbf{x},t)$.
The associated generating function for the dynamic correlation functions and 
cumulants becomes
\begin{eqnarray} \label{eq:gfc}
&&Z[\tilde j,j] = \Big\langle e^{\int d^dx \! \int dt \, 
[{\tilde j}^*(\mathbf{x},t) \, {\tilde \psi}(\mathbf{x},t) 
+ j^*(\mathbf{x},t) \, \psi(\mathbf{x},t)]} \Big\rangle \\
&&\ = \int \! \mathcal D[i {\tilde \psi}] \int \! D[\psi] \,
e^{- \mathcal{A}[\tilde\psi,\psi] + \int d^dx \! \int dt \,
[{\tilde j}^*(\mathbf{x},t) \, {\tilde \psi}(\mathbf{x},t) 
+ j^*(\mathbf{x},t) \, \psi(\mathbf{x},t)]} \ . \nonumber
\end{eqnarray}
We note that $Z[{\tilde j} = 0,j = 0] = 1$ carries no information, 
in stark contrast to the partition function in thermal equilibrium.

The perturbative expansion proceeds around the Gaussian action ($u' = 0$). 
With the Fourier transform convention 
\begin{equation}
\psi(\mathbf{x},t) = \int \frac{d\omega}{2\pi} \, \frac{d^dq}{(2\pi)^d} \,
e^{i(\mathbf{q}\mathbf{x} - \omega t)} \, \psi(\mathbf{q},\omega) \ , 
\end{equation}
and analogously for the response field $\tilde \psi(\mathbf{x},t)$, the Gaussian
action reads in frequency-momentum space:
\begin{eqnarray} \label{eq:gau}
&&\mathcal{A}_0[\tilde\psi,\psi] = \\
&&\quad \int\frac{d\omega}{2 \pi} \, \frac{d^dq}{(2 \pi)^d} \, 
\Bigl( \tilde\psi^*(\mathbf{q},\omega) , \psi^*(\mathbf{q},\omega) \Bigr) \,
A(\mathbf{q},\omega) \left( \begin{array}{c} \tilde\psi(\mathbf{q},\omega) \\
\psi(\mathbf{q},\omega) \end{array} \right) \, , \nonumber
\end{eqnarray}
with the Hermitean harmonic coupling matrix 
\begin{equation} 
A(\mathbf{q},\omega) = \left( \begin{array}{cc} - \gamma / 2 &  
- i \omega + D R(\mathbf{q}) \\ i \omega + D R^*(\mathbf{q}) & 0 \end{array} 
\right) \, ,
\end{equation}
where $R(\mathbf{q}) = r + i r' + \left( 1+ i r_K \right) \mathbf{q}^2$.
Inversion of the $2 \times 2$ matrix yields the bare advanced and retarded 
response propagators as well as the correlation propagator; explicitly, these 
read: 
\begin{eqnarray} \label{eq:prr}
G^0_{\tilde\psi \psi^*}(\mathbf{q},\omega) &=& \frac{1}{i \omega + D \left[ 
r - i r' + \left( 1 - i r_K \right) \mathbf{q}^2 \right]} \ , \nonumber \\
G^0_{\tilde\psi^* \psi}(\mathbf{q},\omega) &=& \frac{1}{- i \omega 
+ D \left[ r + i r' + \left( 1 + i r_K \right) \mathbf{q}^2 \right]} \\
&=& G^{0 \, *}_{\tilde\psi \psi^*}(\mathbf{q},\omega) \ , \nonumber \\
G^0_{\psi^* \psi} (\mathbf{q},\omega) &=& \frac{\gamma}{2} \,
G^0_{\tilde\psi \psi^*}(\mathbf{q},\omega) \, 
G^0_{\tilde\psi^* \psi}(\mathbf{q},\omega) = \frac{\gamma}{2} \, 
\big| G^0_{\tilde\psi^* \psi}(\mathbf{q},\omega) \big|^2 \ . \nonumber
\end{eqnarray}
These expressions can be written in scaling form,
\begin{eqnarray} \label{eq:sc1}
&&G^0_{\tilde\psi^* \psi}(\mathbf{q},\omega)^{-1} = D \, \mathbf{q}^2 
\left( 1 + i r_K + \frac{r + i r'}{\mathbf{q}^2} 
- \frac{i \omega}{D \, \mathbf{q}^2} \right) \, , \\
&&G^0_{\psi^* \psi}(\mathbf{q},\omega) = \frac{\gamma}{2 D^2 \, \mathbf{q}^4} 
\left[ \left( 1 + \frac{r}{\mathbf{q}^2} \right)^2 + \left( 
r_K + \frac{r'}{\mathbf{q}^2} - \frac{\omega}{D \, \mathbf{q}^2} \right)^2 
\right]^{-1} . \nonumber 
\end{eqnarray}
One may set up the diagrammatic perturbation expansion either with these three
propopagators, or equivalently just with the response propagators 
\eqref{eq:prr} and the two-point noise vertex 
$\Gamma^0_{\tilde\psi^* \tilde\psi} = \gamma / 2$, in addition to the 
nonlinear four-point vertices $- \tfrac12 
\Gamma^0_{\tilde\psi^* \psi \psi \psi^*} = - D \, \frac{u'}{6} (1 - i r_U)$ 
and $- \tfrac12 \Gamma^0_{\tilde\psi \psi^* \psi^* \psi} = - D \, \frac{u'}{6} 
(1 + i r_U)$ (computed at symmetrized incoming external wavevectors).
The graphical representations for these elements of the perturbation series
are depicted and explained in Fig.~\ref{fg:fig1}.

\subsection{Relationship with equilibrium critical dynamics models}

It is instructive to rewrite the stochastic differential equation 
\eqref{eq:gpe2} in terms of the coupled real fields $S_1 = \Re \psi$ and 
$S_2 = \Im \psi$, collected into a two-component vector field 
${\vec S}(\mathbf{x},t)$:
\begin{eqnarray} \label{eq:rgpe}
&&\partial_t S_\alpha(\mathbf{x},t) = - D \, \biggl[ \biggl( r - \nabla^2 
+ \frac{u'}{6} \, {\vec S}(\mathbf{x},t)^2 \biggr) \, S_\alpha(\mathbf{x},t) \\
&&\quad - \sum_\beta \epsilon_{\alpha \beta} \, \biggl( r' - r_K \nabla^2 
+ r_U \frac{u'}{6} \, {\vec S}(\mathbf{x},t)^2 \biggr) \, S_\beta(\mathbf{x},t) 
\biggr] + \eta_\alpha(\mathbf{x},t) \nonumber \\
&&\qquad\qquad = F_\alpha^{\rm rel}[{\vec S}](\mathbf{x},t) 
+ F_\alpha^{\rm rev}[{\vec S}](\mathbf{x},t) + \eta_\alpha(\mathbf{x},t) \ ,
\label{eq:rgpe2}
\end{eqnarray}
where $\alpha,\beta = 1,2$, and $\epsilon_{\alpha \beta}$ represents the 
antisymmetric unit tensor in two dimensions (i.e., 
$\epsilon_{12} = - \epsilon_{21} = 1$, $\epsilon_{11} = \epsilon_{22} = 0$). 
The noise correlators \eqref{eq:noi} imply for $\eta_1 = \Re \xi$ and 
$\eta_2 = \Im \xi$:
\begin{eqnarray} \label{eq:rnoi}
&&\left\langle \eta_\alpha(\mathbf{x},t) \right\rangle = 0 \ , \nonumber \\
&&\left\langle \eta_\alpha(\mathbf{x},t) \, \eta_\beta(\mathbf{x'},t') 
\right\rangle = \frac{\gamma}{2} \, \delta_{\alpha \beta} \,
\delta(\mathbf{x} - \mathbf{x}') \, \delta(t - t') \ .
\end{eqnarray}

In Eq. \eqref{eq:rgpe2}, the systematic forces in the Langevin equations have
been decomposed into the dissipative, relaxational term
\begin{equation} \label{eq:rel}
F_\alpha^{\rm rel}[{\vec S}](\mathbf{x},t) = - D \, 
\frac{\delta H[{\vec S}]}{\delta S_\alpha(\mathbf{x},t)} \ ,
\end{equation}
with the standard $O(2)$-symmetric Ginzburg-Landau-Wilson Hamiltonian
\begin{equation} \label{eq:glw}
 H[{\vec S}] = \int \! d^dx \, \biggl[ \frac{r}{2} \, {\vec S}(\mathbf{x})^2 
+ \frac12 \left[ \nabla {\vec S}(\mathbf{x}) \right]^2 
+ \frac{u'}{4 !} \, {\vec S}(\mathbf{x})^4 \biggr] \ ,
\end{equation}
and the reversible contribution
\begin{equation} \label{eq:rev}
F_\alpha^{\rm rev}[{\vec S}](\mathbf{x},t) 
= D \sum_\beta \epsilon_{\alpha \beta} \,
\frac{\delta H'[{\vec S}]}{\delta S_\beta(\mathbf{x},t)} \ ,
\end{equation}
with a second Ginzburg-Landau-Wilson Hamiltonian
\begin{equation} \label{eq:glw2}
 H'[{\vec S}] = \int \! d^dx \, \biggl[ \frac{r'}{2} \, {\vec S}(\mathbf{x})^2 
+ \frac{r_K}{2} \left[ \nabla {\vec S}(\mathbf{x}) \right]^2 
+ r_U \frac{u'}{4 !} \, {\vec S}(\mathbf{x})^4 \biggr] \ .
\end{equation}
In the mean-field approximation, we merely need to simultaneously minimize both
$H$ and $H'$ to obtain possible stationary configurations.
For the temperature-like control parameter $r > 0$ (net gain $\chi < 0$), the 
only homogeneous state is ${\vec S} = 0$, or $\psi = 0$, describing the 
disordered phase, i.e., the absence of a Bose-Einstein condensate.
For $r < 0$ ($\chi > 0$), on the other hand, we encounter the ordered phase 
with finite condensate fraction, namely from minimizing $H$ with a constant 
$|{\vec S}| = \sqrt{6 |r| / u'} = \sqrt{\chi / \kappa} = |\psi|$.
Minimizing the effective Hamiltonian $H'$ yields the second condition 
$|{\vec S}| = \sqrt{6 |r'| / r_U u'} = \sqrt{\mu / \lambda} = |\psi|$, whence
consistency requires that indeed $r' = r_U r$.
The chemical potential then adjusts itself to $\mu = \lambda |\psi|^2$.
In effect, this leaves $r$ as the sole control parameter for the condensation
transition.

We may now consider the following two special cases: 
(i) For parameters $r' = r_K = r_U = 0$, $H'$ vanishes, whence we recover the
$O(2)$-symmetric Model A for purely relaxational critical dynamics towards 
thermal equilibrium, if we impose Einstein's relation (or rescale the fields 
appropriately) 
\begin{equation} \label{eq:fdt}
\gamma = 4 D \, k_{\rm B} T 
\end{equation}
with an {\em effective temperature} $T$.
We shall later employ this parametrization also in a strictly non-equilibrium
setting (with Boltzmann's constant set to $k_{\rm B} = 1$).
A distinct scaling behavior of the noise strength $\gamma$ and the relaxation
rate $D$, and hence the ``temperature'' $T$ indicate a violation of detailed
balance.

(ii) For $r' = r_U r$, $r_K = r_U \not= 0$, $H' = r_K H$, one arrives at an
effective equilibrium dynamics with reversible term 
\begin{equation} \label{eq:rev2}
F_\alpha^{\rm rev}[{\vec S}](\mathbf{x},t) = 
D r_K \sum_\beta \epsilon_{\alpha \beta} \, 
\frac{\delta H[{\vec S}]}{\delta S_\beta(\mathbf{x},t)} \ .
\end{equation}
Its antisymmetry ensures that the associated reversible probability current 
remains divergence-free in the space of dynamical variables $S_\alpha$.
This special situation has been analyzed in Ref.~\onlinecite{DeDominicis75}.

Intriguingly, this kinetics resembles the critical dynamics of Model E for a
non-conserved two-component order parameter field (e.g., the in-plane 
magnetization fluctuations for an XY ferromagnet), reversibly coupled to a
conserved scalar field $M$ (corresponding to the z-component of the 
magnetization in a planar ferromagnet; c.f. 
Refs.~\onlinecite{hohenberg77:_theor,tauberbook}).
Here, however,  $M \propto D r_K$ is spatially uniform and stationary. 
We remark in passing that the dynamic critical exponent of the equilibrium 
Model E is fixed by the fact that the conserved field $M$ generates rotations 
in order parameter space.
Under the assumption of strong dynamic scaling, i.e., proportional divergent
time scales for the critical modes and the conserved quantity, one obtains
$z = d / 2$ exactly in dimensions $d \leq 4$.
Indeed, this constitutes a crucial difference between the equilibrium and the 
driven models: 
The uniform magnetization in the driven case does not scale, whereas the slowly 
varying magnetization field in the equilibrium case does. 
This distinction causes the dynamic critical exponent in the driven and 
equilibrium cases to differ markedly.

Adding an external field term to the Hamiltonian, $H[{\vec h}] = 
H[{\vec S}] - \int \! d^dx \, {\vec h}(\mathbf{x}) \cdot {\vec S}(\mathbf{x})$, 
yields in this special case (ii) the dynamic susceptibilities
\begin{eqnarray} \label{eq:sus}
&&\chi_{\alpha \beta}(\mathbf{x} - \mathbf{x'}, t - t') 
= \frac{\delta \left\langle S_\alpha(\mathbf{x},t) \right\rangle}
{\delta h_\beta(\mathbf{x'},t')} \\
&&\quad = D \, \Bigl\langle S_\alpha(\mathbf{x},t) \, \Bigl[ 
{\tilde S}_\beta(\mathbf{x'},t') + r_K \sum_\gamma \epsilon_{\beta \gamma} \,
{\tilde S}_\gamma(\mathbf{x'},t') \Bigr] \Bigr\rangle \ , \nonumber
\end{eqnarray}
where ${\tilde S}_{1/2}$ represent the (real) Martin-Siggia-Rose response 
fields associated with $S_{1/2}$.
For the original complex fields, these components combine to
\begin{eqnarray} \label{eq:suc}
&&\chi(\mathbf{x} - \mathbf{x'}, t - t') 
= \chi_{11} + \chi_{22} - i \left( \chi_{12} - \chi_{21} \right) \nonumber \\
&&\quad = D \left( 1 + i r_K \right) \left\langle \psi(\mathbf{x},t) \,
{\tilde \psi}^*(\mathbf{x'},t') \right\rangle \ ,
\end{eqnarray}
since the effective Onsager coefficient is $D \left( 1 + i r_K \right)$.
As the system is in thermal equilibrium, the fluctuation-dissipation theorem 
relates the dynamic response with the correlation function 
\begin{equation} \label{eq:dcf}
C(\mathbf{x} - \mathbf{x'}, t - t') = \left\langle \psi^*(\mathbf{x},t) \,
\psi(\mathbf{x'},t') \right\rangle
\end{equation}
through
\begin{equation} \label{eq:fds}
k_{\rm B} T \, \chi(\mathbf{x},t) = - \Theta(t) \, \partial_t C(\mathbf{x},t) 
\ , 
\end{equation}
or equivalently in Fourier space
\begin{equation} \label{eq:fdf}
C(\mathbf{q},\omega) = \frac{2 k_{\rm B} T}{\omega} \, 
{\rm Im} \, \chi(\mathbf{q},\omega) \ .
\end{equation}

For both special situations (i) and (ii), and with Eq.~\eqref{eq:fdt} the 
scaling forms \eqref{eq:sc1} for the retarded response propagator and dynamical 
correlation functions reduce to
\begin{eqnarray} \label{eq:sc2}
&&G^0_{\tilde\psi^* \psi}(\mathbf{q},\omega)^{-1} = D \, \mathbf{q}^2 
\left( 1 + i r_K \right) \left( 1 + \frac{r}{\mathbf{q}^2} 
- \frac{i \omega}{D \, \mathbf{q}^2 \left( 1 + i r_K \right)} \right) \, , 
\nonumber \\
&&G^0_{\psi^* \psi}(\mathbf{q},\omega) = \frac{ 2 k_{\rm B} T}
{D \, \mathbf{q}^4 \left( 1 + r / \mathbf{q}^2 \right)^2} \\ 
&&\qquad\qquad\quad\ \times \left[ 1 + \left( r_K - 
\frac{\omega}{D \, \mathbf{q}^2 \left( 1 + r / \mathbf{q}^2 \right)} \right)^2 
\right]^{-1} . \nonumber 
\end{eqnarray}
With Eqs.~\eqref{eq:suc} and \eqref{eq:dcf}, i.e., 
$\chi(\mathbf{q},\omega) = D \left( 1 + i r_K \right) 
G_{\tilde\psi^* \psi}(\mathbf{q},\omega)$ and 
$C(\mathbf{q},\omega) = G_{\psi^* \psi}(\mathbf{q},\omega)$, 
Eqs.~\eqref{eq:sc2} satisfy the fluctuation-dissipation theorem \eqref{eq:fdf}.
Fourier transformation to the time domain yields explicitly
\begin{eqnarray} \label{eq:sct}
&&G^0_{\tilde\psi^* \psi}(\mathbf{q},t) 
= \Theta(t) \, e^{- D \left( 1 + i r_K \right) \left( r + q^2 \right) t} \ , \\
&&G^0_{\psi^* \psi}(\mathbf{q},t) = \frac{k_{\rm B} T}{r + q^2} \, 
e^{- D \left( 1 + i r_K \right) \left(r + q^2 \right) |t|} \ , \nonumber
\end{eqnarray}
which likewise fulfill Eq.~\eqref{eq:fds}.

\section{Renormalization and Critical Exponents}
\label{sec:renor}

\subsection{Renormalization scheme for ultraviolet divergences}

The perturbation expansion is most conveniently carried out for the 
one-particle irreducible vertex functions, since redundancies are thus 
eliminated in the calculations.
The generating functional for the vertex functions is related to its 
counterpart \eqref{eq:gfc} for the connected correlation functions (cumulants) 
in the standard manner through a Legendre transformation 
\cite{amitbook,itzyksonbook,zinnjustinbook,tauberbook}. 
Here, we merely list the explicit relationships between the two-point vertex
functions and cumulants in Fourier space,
\begin{equation} \label{eq:ga11}
\Gamma_{\tilde\psi \psi^*}(\mathbf{q},\omega) 
= G_{\tilde\psi^* \psi}(\mathbf{q},\omega)^{-1} 
= G^0_{\tilde\psi^* \psi}(\mathbf{q},\omega)^{-1} 
- \Sigma_{\tilde\psi^* \psi}(\mathbf{q},\omega) \ ,
\end{equation}
where the second expression originates from Dyson's equation with the 
associated self-energy $\Sigma$; furthermore
\begin{equation} \label{eq:ga20}
\Gamma_{\tilde\psi^* \tilde\psi}(\mathbf{q},\omega) 
= - \, \frac{G_{\psi^* \psi}(\mathbf{q},\omega)}
{G_{\tilde\psi^* \psi}(\mathbf{q},\omega) \, 
G_{\tilde\psi \psi^*}(\mathbf{q},\omega)} \ ,
\end{equation}
and similarly for the four-point functions, etc.

In the field-theoreticalversion of the renormalization group approach to 
critical phenomena \cite{amitbook,itzyksonbook,zinnjustinbook,kleinert-book,
vasilievbook,folk06,tauberbook}, 
one sends all ultraviolet (UV) cutoffs originating from the short-distance 
physics to infinity.
At and above the upper critical dimension (here, $d_c = 4$) UV 
divergences appear in the perturbation expansion that are absorbed into 
appropriately defined renormalized parameters.
In the vicinity of an RG fixed point, where scale invariance ensues, one may 
then infer the desired infrared (IR) scaling properties of the theory from its 
UV behavior, which is perturbatively accessible, provided one ensures to work 
outside the IR-singular critical region (which here is defined by $r, r' \to 0$ in 
the unrenormalized theory along with $\mathbf{q}, \omega \to 0$).
The field theory action \eqref{eq:act} entails UV divergences for the vertex
functions $\Gamma_{\tilde\psi \psi^*}$, $\Gamma_{\tilde\psi \tilde\psi^*}$, 
and $\Gamma_{\tilde\psi \psi^* \psi^* \psi}$.
The propagator self-energy $\Sigma_{\tilde\psi^* \psi}$ contains quadratic UV
divergences (at $d_c = 4$) that first need to be additively renormalized.
Physically, this corresponds to a fluctuation-induced downward shift of the
critical point (pump rate), $\tau = r - r_c$, and of course the subsequent 
characterization of the critical behavior needs to address the vicinity of the
true phase transition point at $\tau = 0$; i.e., $r_c$ is determined from the
condition $\Gamma_{\tilde\psi \psi^*}(\mathbf{q}=0,\omega=0) = 0$ at $r = r_c$.

The remaining logarithmic UV divergences are then multiplicatively absorbed 
into renormalization factors for which we choose the following conventions:
We define the renormalized counterpart to the two-point vertex function 
\eqref{eq:ga11} as
\begin{equation} \label{eq:zfd}
\Gamma_{\tilde\psi \psi^*}^R(\mathbf{q},\omega) 
= Z \, \Gamma_{\tilde\psi \psi^*}(\mathbf{q},\omega) .
\end{equation}
The complex renormalization constant $Z$ then follows from the singular part 
of its frequency derivative, 
\begin{equation} \label{eq:zfc}
Z^{-1} = i \, \partial_\omega \Gamma_{\tilde\psi \psi^*}(\mathbf{q}=0,\omega) 
|_{\omega=0}^{\rm sing.} \ , 
\end{equation}
evaluated at the normalization point $\tau = \mu^2$ with arbitrary momentum
scale $\mu$, but manifestly outside the IR-singular critical regime.
Next we introduce dimensionless renormalized counterparts to the parameters 
defined in Eqs.~\eqref{eq:rat} and \eqref{eq:fdt} via 
multiplicative renormalization with real $Z$ factors:
\begin{eqnarray} \label{eq:zpd}
&&\tau_R = Z_\tau \tau \, \mu^{-2} \ , \quad u'_R = Z_{u'} u' A_d \, \mu^{d-4} 
\ , \quad T_R = Z_T T \ , \nonumber \\
&&D_R = Z_D D \;\ , \quad r_{K \, R} = Z_{r_K} r_K \;\ , \quad
r_{U \, R} = Z_{r_U} r_U \;\ .
\end{eqnarray}
Here, $A_d = \Gamma(3 - d/2) / 2^{d-1} \pi^{d/2}$ denotes a geometric factor; 
$\Gamma(x)$ indicates Euler's Gamma function.

Note that independent  renormalization constants $Z_T$, $Z_{r_K}$,
and $Z_{r_U}$ only arise in a genuine non-equilibrium setting; the equilibrium 
theory can be fully renormalized through  a {\em real} $Z$ along 
with $Z_\tau$, $Z_D$, and $Z_{u'}$. 
Indeed, the fluctuation-dissipation theorem \eqref{eq:fds} or \eqref{eq:fdf} 
in conjunction with \eqref{eq:ga20} and the definitions \eqref{eq:zfd}, 
\eqref{eq:zpd} implies that 
\begin{equation} \label{eq:eqr}
Z^{-1} = Z_D Z_T
\end{equation}
must hold in thermal equilibrium.
An independent $Z_T$ factor means that the effective ``temperature'' becomes 
scale-dependent in the driven non-equilibrium system.
In contrast, the relation \eqref{eq:eqr} reflects the partition invariance of 
temperature in the renormalization group language: 
all arbitrary system partitions must be in equilibrium with each other. 
Its origin can be traced back to a specific equilibrium symmetry of the 
response functional \cite{tauberbook,sieberer13_long}. 

The renormalization constants will be determined perturbatively to lowest 
non-trivial order in the non-linear coupling $u'$, namely first $Z_D$ and 
$Z_{r_K}$ from $\partial_{q^2} \Gamma_{\tilde\psi \psi^*}(\mathbf{q},\omega=0)
 |_{\mathbf{q}=0}^{\rm sing.}$; subsequently $Z_\tau$ and $Z_{r_U}$ from
$\partial_\tau \Gamma_{\tilde\psi \psi^*}(\mathbf{q}=0,\omega=0)|^{\rm sing.}$,
$Z_T$ from 
$\Gamma_{\tilde\psi \tilde\psi^*}(\mathbf{q}=0,\omega=0)|^{\rm sing.}$, and 
finally $Z_{u'}$ and again $Z_{r_U}$ (as an independent check) from 
$\Gamma_{\tilde\psi \psi^* \psi^* \psi}(\{ \mathbf{q}_i=0 \}, \{ \omega_i=0 \})
 |^{\rm sing.}$.
We shall employ dimensional regularization to compute the associated 
wavevector integrals, whence logarithmic divergences formally appear as 
simple poles in $\epsilon = 4 - d$.
We apply the convenient minimal subtraction scheme, whereupon only these 
$1 / \epsilon$ poles and their residua are incorporated into the $Z$ factors
\cite{amitbook,zinnjustinbook}.

\subsection{Renormalization group equation and RG flow functions}

The renormalization group equation exploits the fact that the unrenormalized
quantities do not depend on the arbitrary momentum renormalization scale $\mu$.
Translated to renormalized correlation or vertex functions, it relates their
properties at different momentum (or length, time) scales, and thus provides 
the desired link between the theory in the ultraviolet, where perturbative 
computations can safely be carried out, and the physically interesting 
infrared region governed by non-trivial critical singularities 
\cite{amitbook,itzyksonbook,zinnjustinbook,kleinert-book,tauberbook}.
Denoting the set of model parameters as 
$\{ p \} = \{ D, \tau, T, r_K, r_U \}$, and introducing  
$u = u' T$, which turns out to be the proper effective 
non-linear coupling in the perturbation series, one obtains for example for 
the two-point vertex function
\begin{eqnarray} \label{eq:rge}
0\!&=&\! \mu \partial_\mu \Gamma_{\tilde\psi \psi^*}(\mathbf{q}, \omega,
\{ p \},u) = \mu \partial_\mu \Bigl[ Z^{-1} 
\Gamma^R_{\tilde\psi \psi^*}(\mathbf{q},\omega,\{ p_R \},u_R) \Bigr] \qquad \\
&=& \biggl( \mu \partial_\mu - \zeta + \sum_p \gamma_p p_R \partial_{p_R} 
+ \beta_u \partial_{u_R} \biggr) \,
\Gamma^R_{\tilde\psi \psi^*}(\mathbf{q},\omega,\{ p_R \},u_R) \ . \nonumber
\end{eqnarray}
Here we have defined Wilson's flow functions 
\begin{equation} \label{eq:wil}
\zeta = \mu \partial_\mu \ln Z \ , \quad 
\gamma_p = \mu \partial_\mu \ln \left( p_R / p \right) \ ,
\end{equation}
-- note that $\zeta$ is complex -- and the RG beta function
\begin{equation} \label{eq:bet}
\beta_u = \mu \partial_\mu u_R 
= u_R \left[ d - 4 + \mu \partial_\mu \ln \left( Z_{u'} Z_T \right) \right] \ .
\end{equation}

The partial differential equation \eqref{eq:rge} is readily solved by the
method of characteristics $\mu \to \mu \ell$, which leads to decoupled 
first-order ordinary differential flow equations for the running parameters 
and the coupling
\begin{eqnarray} \label{eq:run}
&&\ell \frac{d {\tilde p}(\ell)}{d \ell} = {\tilde p}(\ell) \, \gamma_p(\ell) 
\ , \quad {\tilde p}(1) = p_R \ , \nonumber \\
&&\ell \frac{d {\tilde u}(\ell)}{d \ell} = \beta_u(\ell) \ , \qquad\quad 
{\tilde u}(1) = u_R \ .
\end{eqnarray}
The infrared limit is attained as $\ell \to 0$.
Near an infrared-stable RG fixed point given by the zero of the beta function
\eqref{eq:bet}, i.e., $\beta_u(u^*) = 0$ with 
$\partial_{u_R} \, \beta_u |_{u^*} > 0$, the model becomes scale-invariant.
The solutions of the flow equations for the running couplings then become 
simple power laws ${\tilde p}(\ell) \approx p_R \ell^{\gamma_P^*}$, with the 
anomalous scaling dimensions $\gamma_P^* = \gamma_P(u^*)$.
Since all perturbative contributions to the vertex and correlations functions 
consist of integrals over products of Gaussian propagators and vertices, we
observe that the renormalized two-point function takes the form 
$\Gamma^R_{\tilde\psi \psi^*}(\mathbf{q},\omega,\{ p_R \},u_R) = D_R \mu^2 \,
{\widetilde \Gamma} \left( \mathbf{q}^2 \, (1 + i r_{K R}) / \mu^2,
\omega / (D_R \mu^2), \tau_R \, (1 + i r_{U R}), u_R \right)$, c.f.
Eq.~\eqref{eq:sc1}. 
One thus finally arrives at the asymptotic solution of the RG equation 
\eqref{eq:rge} near a fixed point $u^*$, 
\begin{eqnarray} \label{eq:rg1}
&&\Gamma^R_{\tilde\psi \psi^*}(\mathbf{q},\omega,\{ p_R \},u_R) \approx 
D_R \mu^2 \ell^{2 - \zeta(u^*) + \gamma_D^*} \times \\
&&\, {\widetilde \Gamma}\left( 
\frac{\mathbf{q}^2 \, (1 + i r_{K R} \ell^{\gamma_{r_K}^*})}{\mu^2 \ell^2}, 
\frac{\omega}{D_R \mu^2 \ell^{2 + \gamma_D^*}}, \tau_R \ell^{\gamma_\tau^*} \,
(1 + i r_{U R} \ell^{\gamma_{r_U}^*}), u^* \right) \, . \nonumber
\end{eqnarray}
One may similarly proceed for any other vertex function, or, e.g., the 
dynamical correlation function 
\begin{eqnarray} \label{eq:rg2}
&&G^R_{\psi^* \psi}(\mathbf{q},\omega,\{ p_R \},u_R) \approx 
\frac{T_R}{D_R \mu^4} \, 
\ell^{- 4 + 2 \Re \zeta(u^*) + \gamma_T^* - \gamma_D^*} \\
&&\ \times \, {\widetilde C}\left( \frac{\mathbf{q}^2}{\mu^2 \ell^2}, 
\frac{\omega}{D_R \mu^2 \ell^{2 + \gamma_D^*}}, \tau_R \ell^{\gamma_\tau^*},
r_{K R} \ell^{\gamma_{r_K}^*}, r_{U R} \ell^{\gamma_{r_U}^*}, u^* \right) \, .
\nonumber
\end{eqnarray}

\subsection{Renormalization and scaling to one-loop order}

\begin{figure}[t]
\includegraphics[width=0.7\columnwidth]{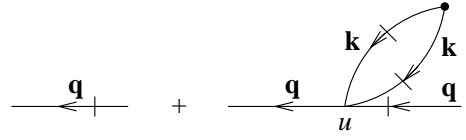}
\caption{One-loop propagator renormalization. 
The loop diagram depicts the first-order correction to the self-energy or
two-point vertex function $\Gamma_{\tilde\psi \psi^*}(\mathbf{q},\omega)$.}
\label{fg:fig2} 
\end{figure}
We now carry out the explicit perturbational analysis of the fluctuation
corrections to first order in $u$, represented through Feynman diagrams with a
single closed propagator loop.
For the two-point vertex function 
$\Gamma_{\tilde\psi \psi^*}(\mathbf{q},\omega)$, the corresponding 
one-particle irreducible graphs are shown in Fig.~\ref{fg:fig2}.
The loop diagram represents the lowest-order contribution to the associated
self-energy $- \Sigma_{\tilde\psi^* \psi}(\mathbf{q},\omega)$, see 
Eq.~\eqref{eq:ga11}.
The ensuing analytic expression is explicitly
\begin{eqnarray} \label{eq:g1l1}
&&\Gamma_{\tilde\psi \psi^*}(\mathbf{q},\omega) \approx - i \omega 
+ D \left[ r + i r' + \left( 1 + i r_K \right) \mathbf{q}^2 \right] 
- \tfrac13 \, D \gamma u' \times \nonumber \\
&&\;\ \left( 1 + i r_U \right) \int_{-\infty}^\infty \frac{d \nu}{2 \pi} 
\int \! \frac{d^dk}{(2 \pi)^d}
\frac{1}{- i \nu + D \left[ r + i r' + (1 + i r_K) \, \mathbf{k}^2 \right]} 
\nonumber \\
&&\qquad\qquad\qquad\qquad \times \,
\frac{1}{i \nu + D \left[ r - i r' + (1 - i r_K) \, \mathbf{k}^2 \right]} \ .
\end{eqnarray}
Upon performing the internal frequency integral via Cauchy's theorem, the 
integrand simplifies considerably.
Setting $\gamma = 4 D T$ (henceforth we employ units where 
Boltzmann's constant $k_{\rm B} = 1$), $r' = r_U r$ and $u = u' T$, one 
arrives at
\begin{eqnarray} \label{eq:gal1}
&&\Gamma_{\tilde\psi \psi^*}(\mathbf{q},\omega) = - i \omega + D \, \Biggl[ 
r \left( 1 + i r_U \right) + \left( 1 + i r_K \right) \mathbf{q}^2 \nonumber \\
&&\qquad\qquad\qquad + \tfrac23 \, u \left( 1 + i r_U \right) 
\int_k \frac{1}{r + k^2} + O(u^2) \Biggr] \, , \quad
\end{eqnarray}
where with $S_d = 1 / 2^{d-1} \pi^{d/2} \, \Gamma(d/2)$:
\begin{equation} \label{eq:wvi}
\int_k f\left( k^2 \right) = \int \! \frac{d^dk}{(2 \pi)^d} \, f\left( 
\mathbf{k}^2 \right) = S_d \int_0^\infty dk \, k^{d-1} f\left( k^2 \right) \, .
\end{equation}

\begin{figure*}[t]
\includegraphics[width=1.6\columnwidth]{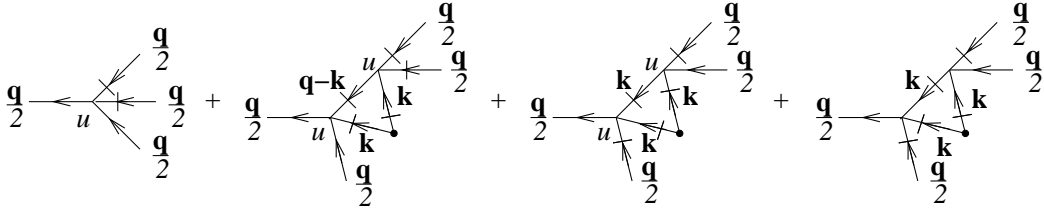}
\caption{One-loop renormalization of the relaxation vertex:
One-particle irreducible diagrams contributing to the four-point vertex
function 
$\Gamma_{\tilde\psi \psi^* \psi^* \psi}(\{ \mathbf{q}/2 \},\{ \omega/2 \})$.}
\label{fg:fig3} 
\end{figure*}
As outlined in Sec.~\ref{sec:renor}.A, we first determine the 
fluctuation-induced shift of the critical point (additive 
renormalization) through the criticality condition 
$\Gamma_{\tilde\psi \psi^*}(\mathbf{q}=0,\omega=0) = 0$ at the true critical 
point $r = r_c = O(u)$.
Eq. \eqref{eq:gal1} yields
\begin{equation} \label{eq:tc1}
r_c(u) = - \tfrac23 \, u \int_k \frac{1}{r_c + k^2} + O(u^2) 
\approx - \tfrac23 \, u \int_k \frac{1}{k^2} + O(u^2) \ ,
\end{equation}
which is negative and represents a downward shift of the critical point: 
fluctuations suppress spontaneous long-range order.
By means of \eqref{eq:dr1} in App.~\ref{sec:appen} one arrives at a 
self-consistent equation for $|r_c|$ which to this order is solved by
\begin{equation} \label{eq:ts1}
r_c(u) \approx 
- \Biggl[ \frac{4 A_d \, u}{3 (d - 2) \, (4 - d)} \Biggr]^{2 / (4 - d)} \, .
\end{equation}
Notice that $r_c(u)$ diverges upon approaching the lower critical dimension 
$d_{lc} = 2$, correctly indicating that the critical point is driven towards 
zero and there emerges no truly long-range order with spatially homogeneous 
condensate (in fact, an isotropic driven-dissipative Bose gas in two 
dimensions cannot even support quasi long-range order, see 
Ref.~\onlinecite{altman13}). 
Furthermore, both non-equilibrium parameters $r_K$ and $r_U$ have dropped out:
Eq.~\eqref{eq:tc1} just represents the {\em equilibrium} critical point shift.
Indeed, since the fluctuation loop in Eq.~\eqref{eq:gal1} is proportional to 
the factor $1 + i r_U$, we may define the true distances from the critical 
point $\tau = r - r_c$ and $\tau' = r_U \left( r - r_c \right) = r_U \, \tau$, 
which vanish {\em simultaneously} as $r \to r_c(u)$, even when fluctuation 
effects are included (to first order), thus preserving the basic mean-field 
scenario.

In terms of $\tau$, since $r_c = O(u)$ one can now rewrite Eq.~\eqref{eq:gal1} 
to this order:
\begin{eqnarray} \label{eq:gml1}
&&\Gamma_{\tilde\psi \psi^*}(\mathbf{q},\omega) = - i \omega 
+ D \left( 1 + i r_K \right) \mathbf{q}^2  \nonumber \\
&&\quad + D \, \tau \left( 1 + i r_U \right) \Biggl[ 1 -  \tfrac23 \, u 
\int_k \frac{1}{k^2 \left( \tau + k^2 \right)} \Biggr] +  O(u^2) \ . \quad
\end{eqnarray}
Just as in thermal equilibrium, the UV singularities contained in the 
wavevector integral can be entirely absorbed into a multiplicative 
renormalization of the parameter $\tau$, see Eq.~\eqref{eq:zpd}:
\begin{eqnarray} \label{eq:1zt}
&&Z_\tau = 1 -  \tfrac23 \, u \int_k \frac{1}{k^2 \left( \tau + k^2 \right)} 
= 1 - \frac{4 u A_d \, \mu^{d-4}}{3 (d - 2) (4 - d)} + O(u^2) \nonumber \\
&&\quad \to 1 - \frac{2 u A_d \, \mu^{-\epsilon}}{3 \epsilon} 
+ O(u^2, \epsilon^0) \ , 
\end{eqnarray}
where we have used \eqref{eq:dr2} in App.~A.1, and in the final step 
applied the minimal subtraction scheme, wherein only the $1 / \epsilon$ 
pole and its residuum at the upper critical dimension $d_c$ are included in 
the renormalization constant $Z_\tau$.
Since in addition there exists no one-loop correction to the noise vertex, 
see Fig.~\ref{fg:fig4} below, i.e.,
$\Gamma_{\tilde\psi^* \tilde\psi}(\mathbf{q},\omega) = - 2 D T + O(u^2)$, we 
infer that to this order
\begin{equation} \label{eq:1lz}
Z = Z_T = Z_D = Z_{r_K} = Z_{r_U} = 1 + O(u^2) \ .
\end{equation}
The associated Wilson RG flow functions \eqref{eq:wil} and anomalous 
dimensions all vanish in the one-loop approximation,
\begin{equation} \label{eq:1lg}
 \zeta = \gamma_T = \gamma_D = \gamma_{r_K} = \gamma_{r_U} 
= 0 + O(u_R^2) \ ,
\end{equation}
and the sole non-trivial RG flow function to first order in $u_R$ is
\begin{equation} \label{eq:1gt}
\gamma_\tau = - 2 + \tfrac23 \, u_R + O(u_R^2) \ .
\end{equation}

In order to compute the beta function \eqref{eq:bet} for the non-linear 
coupling $u$ and obtain the RG fixed points, we require the renormalization
of the four-point vertex function $\Gamma_{\tilde\psi \psi^* \psi^* \psi}$.
Its tree and one-loop contributions are depicted in Fig.~\ref{fg:fig3}.  
Carrying out the internal frequency integrals and combining the three loop
contributions, one finally arrives at
\begin{eqnarray} \label{eq:g3l1}
&&\Gamma_{\tilde\psi \psi^* \psi^* \psi}(\{ \mathbf{q}/2 \},\{ \omega/2 \}) 
= D \, \tfrac13 \, u' \left( 1 + i r_U \right) \Biggl[ 1 - \tfrac13 \, u 
\left( 1 + i r_U \right) \nonumber \\
&&\quad\ \times \int_k \frac{1}{r + k^2} \,
\frac{1}{- \frac{i \omega}{2 D} + r + i r' + \left( 1 + i r_K \right) \left( 
\frac{\mathbf{q}^2}{2} - \mathbf{q} \cdot \mathbf{k} + k^2 \right)} \nonumber\\
&&\qquad\qquad\qquad\qquad\quad 
- \tfrac43 \, u \int_k \frac{1}{\left( r + k^2 \right)^2} + O(u^2) \Biggr] \ .
\end{eqnarray}
To this order in $u$, we may replace $r$ with $\tau$ in the integrals, and
$r' = r_U \, r$ with $\tau' = r_U \, \tau$, and evaluate at the normalization
point $\tau = \mu^2$ ($\tau_R = 1$, safely outside the IR-singular region), 
with $\mathbf{q} = 0$, $\omega = 0$.  
By means of Eqs.~\eqref{eq:dr2} and \eqref{eq:dr3}, we obtain in minimal 
subtraction:
\begin{eqnarray} \label{eq:g3s1}
&&\Gamma_{\tilde\psi \psi^* \psi^* \psi}(\{ 0 \}, \{ 0 \}) |^{\rm sing.} 
= D \, \tfrac13 \, u' \left( 1 + i r_U \right) \nonumber \\
&&\quad \times \, \Biggl[ 
1 - \frac{1 + i r_U}{1 + i r_K} \, \frac{u A_d \, \mu^{-\epsilon}}{3 \epsilon} 
- \frac{4 u A_d \, \mu^{-\epsilon}}{3 \epsilon} 
+ O(u^2, \epsilon^0) \Biggr] \ . \qquad
\end{eqnarray}
Separating out the real and imaginary parts yields the renormalization 
constants
\begin{eqnarray} \label{eq:1zu}
&&Z_{u'} = 1 + \frac{(r_U - r_K)^2}{1 + r_K^2} \, 
\frac{u A_d \, \mu^{-\epsilon}}{3 \epsilon} 
- \frac{5 u A_d \, \mu^{-\epsilon}}{3 \epsilon} + O(u^2, \epsilon^0) \ , 
\nonumber \\
&&Z_{r_U} = 1 - \frac{(r_U - r_K) \, (1 + r_U^2)}{r_U \, (1 + r_K^2)} \, 
\frac{u A_d \, \mu^{-\epsilon}}{3 \epsilon} + O(u^2, \epsilon^0) \ . 
\label{eq:1zr}
\end{eqnarray}
Notice that consistency with the one-loop analysis of the propagator 
self-energy that led to $Z_{r_U} = 1 + O(u^2)$ demands that $r_U^* = r_K^*$ at 
the stable RG fixed point.

With $Z_T = 1 + O(u^2)$, the RG beta function \eqref{eq:bet} becomes
\begin{equation} \label{eq:1bu}
\beta_u = u_R \, \Biggl[ - \epsilon + \tfrac53 \, u_R 
- \frac{\Delta_R^2}{3 (1 + r_{K R}^2)} \, u_R + O(u_R^2) \Biggr] \ ,
\end{equation}
where $\Delta_R = r_{U R} - r_{K R}$.
For $d > d_c = 4$ ($\epsilon < 0$), its only stable zero is the Gaussian fixed
point $u_0^* = 0$ (with $\partial_u \, \beta_u |_{u_0^* = 0} = - \epsilon$), 
which implies mean-field scaling exponents $\nu = 1/2$, $\eta = 0$, $z = 2$, 
and also $\eta_c = 0 = \eta'$.
At the upper critical dimension, the RG flow tends to zero only slowly, 
inducing logarithmic corrections to the mean-field power laws.
In dimensions $d < 4$, a non-trivial fixed point emerges:
\begin{equation} \label{eq:1fp}
u^* = \frac{3 (1 + r_{K R}^2)}{5 (1 + r_{K R}^2) - \Delta_R^2} \, \epsilon 
+ O(\epsilon^2) 
\end{equation}
(provided the denominator is positive). 
It depends parametrically on $r_{K R}$ and the difference $\Delta_R$, and leads 
to non-Gaussian critical exponents.
We next study the RG beta function associated with $\Delta_R$.
Since $\gamma_{r_K} = 0$, we obtain from Eq.~\eqref{eq:1zr}
\begin{eqnarray} \label{eq:1bd}
&&\beta_\Delta = \mu \partial_\mu \Delta_R 
= r_{U R} \gamma_{r_U} - r_{K R} \gamma_{r_K} \nonumber \\
&&\quad\ = \Delta_R \left( 1 + \frac{2 r_{K R} \, \Delta_R + \Delta_R^2}
{1 + r_{K R}^2} \right) \frac{u_R}{3} + O(u_R^2) \ .
\end{eqnarray}
At the Gaussian fixed point $u_0^* = 0$, any constant values of $r_{K R}$ and 
$r_{U R}$ are allowed.
For $d < 4$, at the non-trivial, positive, and stable fixed point 
\eqref{eq:1fp}, the only real zero of Eq.~\eqref{eq:1bd} is indeed 
$\Delta^* = 0$, whence, as anticipated, $r_U^* = r_K^*$, and the RG fixed point
becomes independent of $r_K$:
\begin{equation} \label{eq:efp}
\Delta^* = 0 \ , \quad u^* = \tfrac35 \, \epsilon + O(\epsilon^2) \ .
\end{equation}
This is just the equilibrium XY model fixed point for the $O(2)$-symmetric 
Ginzburg-Landau-Wilson Hamiltonian.
We note that the stability matrix eigenvalues at the infrared-stable RG fixed 
point \eqref{eq:efp} are 
$\partial_{u_R} \, \beta_u |_{\Delta^* = 0, u^*} = \epsilon$ and 
$\partial_{\Delta_R} \, \beta_\Delta |_{\Delta^* = 0, u^*} = \epsilon / 5$.
Therefore the RG flow will typically first approach the non-equilibrium fixed 
line \eqref{eq:1fp}, and subsequently tend towards the equilibrium fixed point 
$\Delta_R \to 0$ along this critical surface.
 At this point, the system has already become effectively
{\em thermalized}, and is described by the special case (ii) discussed in
Sec.~\ref{sec:model}C; thus Eq.~\eqref{eq:eqr} holds, albeit trivially to 
one-loop order, see \eqref{eq:1lz}.

With the one-loop results \eqref{eq:1lg}, the solutions \eqref{eq:rg1} and 
\eqref{eq:rg2} of the RG equations for the inverse response propagator and the 
dynamical correlation function simplify drastically at the one-loop 
equilibrium fixed point.
According to Eq.~\eqref{eq:sc2} and applying the matching condition 
$\ell = |\mathbf{q}| / \mu$, they reduce to the following scaling laws for the
dynamical susceptibility:
\begin{eqnarray} \label{eq:p1l}
&&\chi^R(\mathbf{q},\omega,\{ p_R \},u_R)^{-1} = D_R^{-1} 
\Gamma^R_{\tilde\psi \psi^*}(\mathbf{q},\omega,\{ p_R \},u_R) \nonumber \\
&&\ \approx |\mathbf{q}|^2 \left( 1 + i r_{K R} \right) 
{\hat \Gamma}\left( \frac{\omega}{D_R \, |\mathbf{q}|^2 \left( 1 + i r_{K R} 
\right)}, \frac{\tau_R}{|\mathbf{q} / \mu|^{1 / \nu}} \right) \, , \qquad
\end{eqnarray}
where we have omitted fixed, constant arguments and identified the inverse 
correlation length exponent
\begin{equation} \label{eq:1nu}
\nu^{-1} = - \gamma_\tau^* = 2 - \tfrac25 \, \epsilon + O(\epsilon^2) \ ,
\end{equation}
and for the dynamical correlation function
\begin{equation} \label{eq:c1l}
G^R_{\psi^* \psi}(\mathbf{q},\omega,\{ p_R \},u_R) \approx 
\frac{T_R}{D_R \, |\mathbf{q}|^4} \, 
{\hat C}\left( \frac{\omega}{D_R \, |\mathbf{q}|^2}, 
\frac{\tau_R }{|\mathbf{q} / \mu|^{1 / \nu}}, r_{K R} \right) \, .
\end{equation}
These expressions imply that $\eta = 0 + O(\epsilon^2) = \eta_c = \eta'$ and 
$z = 2 + O(\epsilon^2)$ to one-loop order.

\subsection{Two-loop analysis and renormalization}

\begin{figure}[t]
\includegraphics[width=0.55\columnwidth]{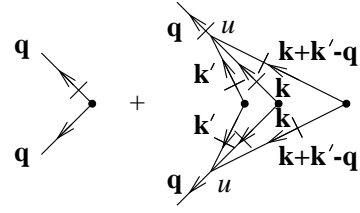}
\caption{Two-loop renormalization of the noise vertex: 
There is no fluctuation correction to first order in $u$; the two-loop graph
represents the lowest-order contribution to the vertex function
$\Gamma_{\tilde\psi^* \tilde\psi}(\mathbf{q},\omega)$.}
\label{fg:fig4} 
\end{figure}
In order to obtain non-trivial dynamic critical and drive exponents, we need to 
proceed to the next order in the perturbational and dimensional $\epsilon$
expansion.
The two-point noise vertex is only renormalized to two-loop order, as shown in
Fig.~\ref{fg:fig4}.
Carrying out both internal frequency integrals associated with the closed
propagator loops, one arrives to second order in the non-linear coupling $u$ at
\begin{eqnarray} \label{eq:n2l}
&&\Gamma_{\tilde\psi^* \tilde\psi}(\mathbf{q},\omega) = - 2 D T \, 
\Biggl[ 1 + \tfrac29 \, u^2 \left( 1 + r_U^2 \right) \nonumber \\
&&\qquad \times \int_k \frac{1}{r + k^2} \int_p \frac{1}{r + p^2} \,
\frac{1}{r + (\mathbf{q} - \mathbf{k} - \mathbf{p})^2} \, \times
\nonumber \\ 
&&\!\!\!\!\!\!\! 
{\rm Re} \, \frac{1}{\frac{- i \omega}{D} + 3 r + i r' + (1 - i r_K) 
(\mathbf{q} - \mathbf{k} - \mathbf{p})^2 + (1 + i r_K) (k^2 + p^2)} \nonumber \\
&&\qquad\qquad\qquad\qquad\qquad\qquad\qquad\qquad + O(u^3) \Biggr] \ .
\end{eqnarray}
Setting $r' = r_U r$, and replacing $r = \tau + O(u)$ in the integrands, we
find
\begin{equation} \label{eq:nr2}
\Gamma_{\tilde\psi^* \tilde\psi}(0,0) =  - 2 D T \, \Bigl[ 1 + 
\tfrac29 \, u^2 \left( 1+ r_U^2 \right) {\rm Re} \, I_d\left( \tau, r_K, r_U 
\right) + O(u^3) \Bigr] \ ,
\end{equation}
with the nested wave vector integral
\begin{eqnarray} \label{eq:idt}
&&I_d\left( \tau, r_K, r_U \right) = \int_k \frac{1}{\tau + k^2} \int_p 
\frac{1}{\tau + p^2} \, \frac{1}{\tau + (\mathbf{k} + \mathbf{p})^2} \nonumber\\
&&\;\ \times \, \frac{1}{2 \tau + (1 + i r_U) \, \tau + 2 (k^2 + p^2) 
+ 2 (1 - i r_K) \, \mathbf{k} \cdot \mathbf{p}} \ . \qquad
\end{eqnarray}
Evaluating this integral at the normalization point $\tau = \mu^2$, and 
isolating its UV divergences in the form of $1 / \epsilon$ poles then yields 
the $Z$ factor product (in minimal subtraction)
\begin{equation} \label{eq:2zt}
Z_D \, Z_T = 1 + \tfrac29 \, u^2 \left( 1 + r_U^2 \right) {\rm Re} \, 
I_d\left( \mu^2, r_K, r_U \right)\big|^{\rm sing.} + O(u^3) \ . 
\end{equation}

\begin{figure*}[t]
\includegraphics[width=1.5\columnwidth]{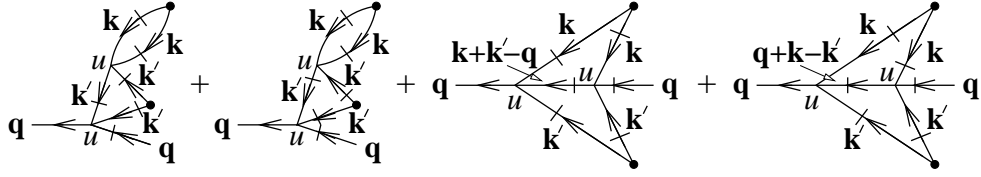}
\caption{Two-loop one-particle-irreducible Feynman diagrams contributing to 
the propagator self-energy or two-point vertex function 
$\Gamma_{\tilde\psi \psi^*}(\mathbf{q},\omega)$.
The two graphs on the right induce non-classical values for the exponents 
$\eta$, $z$, $\eta_c$, and $\eta'$.}
\label{fg:fig5} 
\end{figure*}
The two-loop Feynman graphs contributing to the retarded response propagator 
self-energy or vertex function $\Gamma_{\tilde\psi \psi^*}(\mathbf{q},\omega)$ 
are depicted in Fig.~\ref{fg:fig5}.
The first two closed diagrams (on the left) yield contributions that are 
independent of the external wavevector $\mathbf{q}$ and frequency $\omega$, 
and combine to
\begin{equation} \label{eq:s21}
- D \, \tfrac49 \, u^2 \left( 1 + i r_U \right) \int_k \frac{1}{r + k^2} 
\int_p \frac{1}{\left( r + p^2 \right)^2} \ .
\end{equation}
The two graphs to the right, with the wavevectors distributed as indicated in 
Fig.~\ref{fg:fig5}, yield after internal frequency integration
\begin{eqnarray} \label{eq:s22}
&&- D \, \tfrac29 \, u^2 \left( 1 + i r_U \right) \int_k \frac{1}{r + k^2} 
\int_p \frac{1}{r + p^2} \times \\
&&\!\!\!\!\! \Biggl[ \frac{1 - i r_U}{\frac{- i \omega}{D} + 3 r + i r' + 
(1 - i r_K) (\mathbf{q} - \mathbf{k} - \mathbf{p})^2 + (1 + i r_K) (k^2 + p^2)}
\nonumber \\
&&\!\!\!\!\!\!\! + \frac{2 \, (1 + i r_U)}{\frac{- i \omega}{D} + 3 r + i r' 
+ (1 + i r_K) \bigl[ p^2 + (\mathbf{q} - \mathbf{k} - \mathbf{p})^2 \bigr] 
+ (1 - i r_K) k^2} \Biggr] . \nonumber 
\end{eqnarray}
Symmetrizing with respect to the internal wavevectors 
$\mathbf{k} \leftrightarrow \mathbf{p} \leftrightarrow \mathbf{q} - \mathbf{k} 
- \mathbf{p}$ simplifies this expression markedly, and the sum of 
Eqs.~\eqref{eq:gal1}, \eqref{eq:s21}, and \eqref{eq:s22} can be written as
\begin{eqnarray} \label{eq:gal2}
&&\Gamma_{\tilde\psi \psi^*}(\mathbf{q},\omega) = - i \omega + D \, \Biggl[ 
r \left( 1 + i r_U \right) + \left( 1 + i r_K \right) \mathbf{q}^2 \nonumber \\
&&\quad + \tfrac23 \, u \left( 1 + i r_U \right) \int_k \frac{1}{r + k^2} \ 
\Biggl( 1 - \tfrac23 \, u \int_p \frac{1}{\left( r + p^2 \right)^2} \Biggr)
\nonumber \\
&&- \tfrac29 \, u^2 \left( 1 + i r_U \right) \int_k \frac{1}{r + k^2} \int_p 
\frac{1}{r + p^2} \, \frac{1}{r + (\mathbf{q} - \mathbf{k} - \mathbf{p})^2} \ 
\Biggl( 1 - \nonumber \\
&&\!\! \frac{\frac{- i \omega}{D} - i \, (r_U - r_K) \, 
[(\mathbf{q} - \mathbf{k} - \mathbf{p})^2 - k^2 - p^2]}{\frac{- i \omega}{D} 
+ 3 r + i r' + (1 - i r_K) (\mathbf{q} - \mathbf{k} - \mathbf{p})^2 
+ (1 + i r_K) (k^2 + p^2)} \Biggr) \nonumber \\
&&\qquad\qquad\qquad\qquad\qquad\qquad\qquad\qquad + O(u^3) \Biggr] \ .
\end{eqnarray}

For vanishing external wavevector and frequency, we obtain with $r' = r_U r$:
\begin{eqnarray} \label{eq:g0l2}
&&\frac{\Gamma_{\tilde\psi \psi^*}(0,0)}{D \left( 1 + i r_U \right)} = 
r + \tfrac23 \, u \int_k \frac{1}{r + k^2} \, \Biggl( 1 - \tfrac23 \, u 
\int_p \frac{1}{\left( r + p^2 \right)^2} \Biggr) \nonumber \\
&&\quad - \tfrac29 \, u^2 \int_k \frac{1}{r + k^2} \int_p \frac{1}{r + p^2} \, 
\frac{1}{r + (\mathbf{k} + \mathbf{p})^2} \ \Biggl( 1 + \\
&&\;\ \frac{2 i \, (r_U - r_K) \, \mathbf{k} \cdot \mathbf{p}}{2 r + 
(1 + i r_U) r + 2 (k^2 + p^2) + 2 (1 - i r_K) \, \mathbf{k} \cdot \mathbf{p}} 
\Biggr) + O(u^3) \ . \nonumber
\end{eqnarray}
At the stable RG fixed point \eqref{eq:efp} with $r_U^* = r_K^*$, the 
right-hand side of Eq.~\eqref{eq:g0l2} reduces to the standard two-loop 
additive and multiplicative temperature renormalizations for the mass 
parameter $r$.
The fluctuation-induced $T_c$ shift, as well as the renormalization constant
$Z_\tau$ and hence the correlation length exponent $\nu$, remain identical to 
those of the XY model or $O(2)$-symmetric Model A in thermal equilibrium to 
this order.
The remaining multiplicative renormalization factors follow from the frequency
and wavevector derivatives of Eq.~\eqref{eq:gal2} at the normalization point 
$\tau = r + O(u) = \mu^2$: 
According to Eq.~\eqref{eq:zfc}
\begin{equation} \label{eq:zl2}
Z^{-1} = 1 + \tfrac29 \, u^2 \left( 1 + i r_K \right) 
I_d\left( \mu^2, r_K, r_U = r_K \right)\big|^{\rm sing.} + O(u^3) \ ,
\end{equation}
whence subsequently $Z_D$ and $Z_{r_K}$ can be determined from the singular 
contributions to
\begin{eqnarray} \label{eq:dqg}
&&\partial_{q^2} \Gamma_{\tilde\psi \psi^*}(\mathbf{q},\omega=0) 
|^{\rm sing.}_{\mathbf{q}=0} = Z^{-1} Z_D \, D \left( 1 + i \, Z_{r_K} r_K 
\right) \\
&&\quad = D \left( 1 + i r_K \right) \Bigl[ 1 - \tfrac29 \, u^2 \, 
\partial_{q^2} D_d\left( \mu^2,\mathbf{q} \right)
\big|^{\rm sing.}_{\mathbf{q}=0} + O(u^3) \Bigr] \ , \quad \nonumber 
\end{eqnarray}
where
\begin{equation} \label{eq:dqt}
D_d(\tau,\mathbf{q}) = \int_k \frac{1}{\tau + k^2} \int_p \frac{1}{\tau + p^2} 
\, \frac{1}{\tau + (\mathbf{q} - \mathbf{k} - \mathbf{p})^2} \ .
\end{equation}
The ultraviolet singularities to be captured in $Z_{r_K}$ encode a novel
scaling exponent that describes the weight and fadeout of coherent quantum 
fluctuations relative to their thermal, dissipative counterparts.

Appendix~A.1 details how the UV-singular part is extracted from this nested 
wavevector integral in the form of a simple $1 / \epsilon$ pole and its 
residuum, applying dimensional regularization with minimal subtraction.
Thus, Eq.~\eqref{eq:dqg} yields with Eq.~\eqref{eq:dq4}:
\begin{equation} \label{eq:zf1}
Z^{-1} Z_D \left( 1 + i \, Z_{r_K} r_K \right) = \left( 1 + i r_K \right) 
\Biggl[ 1 + \frac{u^2 A_d^2 \, \mu^{- 2 \epsilon}}{36 \epsilon} 
+ O(u^3, \epsilon^0) \Biggr] \ .
\end{equation}
The evaluation of the integral \eqref{eq:idt}, detailed in Apps.~A.1 and A.2, 
gives
\begin{eqnarray} \label{eq:zf2}
&&{\rm Re} \, Z^{-1} = 1 + \frac{u^2 A_d^2 \, \mu^{- 2 \epsilon}}{36 \epsilon} \,
\Biggl[ \frac{3 + r_K^2}{1 + r_K^2} \ln \frac{16}{9 + r_K^2} \nonumber \\
&&\qquad\ - \frac{1 - r_K^2}{1 + r_K^2} \ln \left( 1 + r_K^2 \right) 
+ \frac{4 r_K}{1 + r_K^2} \left( \arctan r_K - \arctan \frac{r_K}{3} \right)
\nonumber \\
&&\qquad\qquad\qquad\qquad\ 
+ \sqrt{\tfrac32 \left( s_k + 1 \right)} \, L(r_K) \Biggr] + O(u^3) \ , \\
&&{\rm Im} \, Z^{-1} = \frac{u^2 A_d^2 \, \mu^{- 2 \epsilon}}{36 \epsilon} \,
\Biggl[ \frac{2 r_K}{1 + r_K^2} \, \ln \frac{16}{(9 + r_K^2) \, (1 + r_K^2)} 
\nonumber \\
&&\qquad\qquad\qquad\quad - 2 \, \frac{1 - r_K^2}{1 + r_K^2} \arctan r_K 
+ 2 \, \frac{3 + r_K^2}{1 + r_K^2} \arctan \frac{r_K}{3} \nonumber \\
&&\qquad\qquad\qquad\qquad\ + \sqrt{\tfrac32 \left( s_K - 1 \right)} \, L(r_K) 
\Biggr] + O(u^3) \ , 
\label{eq:zf3}
\end{eqnarray}
where $s_k = \sqrt{1 + \tfrac49 \, r_K^2}$ and the logarithmic function 
$L(r_K)$ is given in Eq.~\eqref{eq:lrk}.
Finally, at the infrared-stable fixed point where $r_U^* = r_K^*$:
\begin{eqnarray} \label{eq:zf4}
&&
Z_D \, Z_T = 1 + \frac{u^2 A_d^2 \, \mu^{- 2 \epsilon}}{36 \epsilon} \, 
\Biggl[ 3 \ln \frac{16}{9 + r_K^2} - \ln \left(1 + r_K^2 \right) \nonumber \\
&&\qquad\qquad\qquad\qquad
+ 2 r_K \left( \arctan r_K + \arctan \frac{r_K}{3} \right) \nonumber \\ 
&&\quad + \sqrt{\tfrac32} \left( \sqrt{s_K + 1} + r_K \sqrt{s_K - 1} \, \right) 
L(r_K) \Biggr] + O(u^3) \ . \quad
\end{eqnarray}

Carefully separating the real and imaginary parts in Eq.~\eqref{eq:zf1} and 
inserting Eqs.~\eqref{eq:zf2}, \eqref{eq:zf3} allows us to compute the desired 
renormalization constants to two-loop order:
\begin{eqnarray} \label{eq:zrk}
&&Z_{r_K} = 1 - \frac{u^2 A_d^2 \, \mu^{- 2 \epsilon}}{36 \epsilon} \,
\Biggl[ 2 \ln \frac{16}{(9 + r_K^2) \, (1 + r_K^2)} \nonumber \\
&&\qquad\qquad - 2 \, \frac{1 - r_K^2}{r_K} \arctan r_K 
+ 2 \, \frac{3 + r_K^2}{r_K} \arctan \frac{r_K}{3} \nonumber \\ 
&&\qquad\qquad + \frac{1 + r_K^2}{r_K} \sqrt{\tfrac32 \left( s_K - 1 \right)} 
\, L(r_K) \Biggr] + O(u^3) \ ,
\end{eqnarray}
and
\begin{eqnarray} \label{eq:zd2} 
&&Z_D = 1 - \frac{u^2 A_d^2 \, \mu^{- 2 \epsilon}}{36 \epsilon} \,
\Biggl[ - 1 + \frac{3 - r_K^2}{1 + r_K^2} \ln \frac{16}{9 + r_K^2} \nonumber \\
&&\qquad\qquad\qquad\qquad\ 
- \frac{1 - 3 r_K^2}{1 + r_K^2} \ln \left( 1 + r_K^2 \right) \nonumber \\
&&\qquad\quad + 2 r_K \, \Biggl( \frac{3 - r_K^2}{1 + r_K^2} \arctan r_K 
- \frac{5 + r_K^2}{1 + r_K^2} \arctan \frac{r_K}{3} \Biggr) \nonumber \\
&&\;\ + \sqrt{\tfrac32} \left( \sqrt{s_K + 1} - r_K \sqrt{s_K - 1} \right) 
L(r_K) \Biggr] + O(u^3) \ ; 
\end{eqnarray}
at last, Eq.~\eqref{eq:zf4} gives with Eq.~\eqref{eq:zd2}:
\begin{eqnarray} \label{eq:zt2} 
&&Z_T = 
1 - \frac{u^2 A_d^2 \, \mu^{- 2 \epsilon}}{36 \epsilon} \, \Biggl[ 1 
- 2 \, \frac{3 + r_K^2}{1 + r_K^2} \ln \frac{16}{9 + r_K^2} \nonumber \\
&&\qquad\quad + 2 \, \frac{1 - r_K^2}{1 + r_K^2} \ln \left( 1 + r_K^2 \right) 
- \frac{8 r_K}{1+ r_K^2} \arctan \frac{2 r_K}{3 + r_K^2} \nonumber \\
&&\qquad\qquad\qquad\quad 
- \sqrt{6 \left( s_K + 1 \right)} \, L(r_K) \Biggr] + O(u^3) \ .
\end{eqnarray}

\subsection{Scaling and critical exponents to order $\epsilon^2$}

\begin{figure}[t]
\includegraphics[width=\columnwidth]{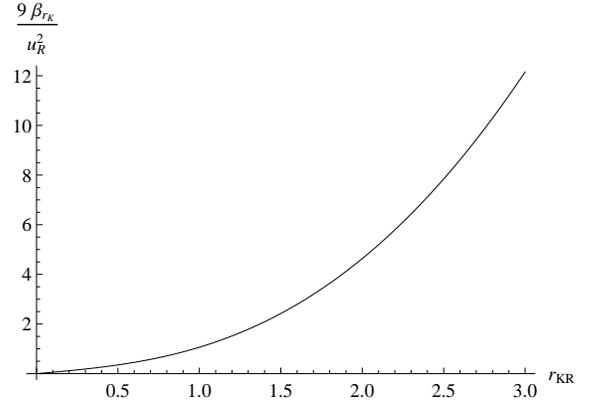}
\caption{Two-loop RG beta function $9 \beta_{R_K} / u_R^2$ as function of the 
renormalized non-equilibrium parameter $r_{K R}$, Eq.~\eqref{eq:brk}.}
\label{fg:fig6} 
\end{figure}

In order to explore possible fixed points, we compute the RG beta function for 
the non-equilibrium parameter $r_K$, using Eqs.~\eqref{eq:zpd} and 
\eqref{eq:zrk}:
\begin{eqnarray} \label{eq:brk}
&&\beta_{r_K} = \mu \partial_\mu r_{K R} = r_{K R} \gamma_{r_K} = \frac{u_R^2}{9}
\, \Biggl[ r_{K R} \ln \frac{16}{(9 + r_{K R}^2) \, (1 + r_{K R}^2)} \nonumber \\
&&\qquad\ + 2 \left( 1 + r_{K R}^2 \right) \arctan r_{K R} - \left( 
3 + r_{K R}^2 \right) \arctan \frac{2 r_{K R}}{3 + r_{K R}^2} \nonumber \\ 
&&\qquad\ + \tfrac12 \left( 1 + r_{K R}^2 \right) 
\sqrt{\tfrac32 \left( s_{K R} - 1 \right)} \, L(r_{K R}) \Biggr] + O(u_R^3) \ .
\end{eqnarray}
At the equilibrium fixed point $r_K^* = 0$, we have 
$\partial_{r_{K R}} \beta_{r_K} |_{r_K^* = 0} = 
\tfrac29 \ln \tfrac43 \, u_R^2 > 0$, whence it is infrared-stable. 
Indeed, as shown in Fig.~\ref{fg:fig6}, $\beta_{r_K}(r_{K R})$ is a 
monotonically growing function of the renormalized non-equilibrium parameter 
$r_{K R}$, and $r_K^* = 0$ is its only zero. 
Setting $r_{K R} = 0$, Wilson's flow functions \eqref{eq:wil}, readily derived 
from Eqs.~\eqref{eq:zf2}, \eqref{eq:zf3}, \eqref{eq:zrk}, \eqref{eq:zd2}, and
\eqref{eq:zt2}, simplify drastically:
\begin{eqnarray} \label{eq:2lw}
&&\zeta = \frac{u_R^2}{18} \, 6 \ln \tfrac43 + O(u_R^3) \ , \nonumber \\
&&\gamma_{r_K} = \frac{u_R^2}{18} \, 4 \ln \tfrac43 + O(u_R^3) \ , \\
&&\gamma_D = \frac{u_R^2}{18} \left( 6 \ln \tfrac43 - 1 \right) + O(u_R^3) \ , 
\nonumber \\ 
&&\gamma_T = - \frac{u_R^2}{18} \left( 12 \ln \tfrac43 - 1 
\right) + O(u_R^3) \ . \nonumber
\end{eqnarray}
Once the system has reached a thermalized state, which here happens already at 
the one-loop level, the exact equilibrium relation \eqref{eq:eqr} enforces the 
identity 
\begin{equation} \label{eq:eqi}
\zeta + \gamma_D + \gamma_T = 0
\end{equation}
with a {\em real} $\zeta$, which is in fact satisfied by the explicit two-loop 
results \eqref{eq:2lw}.

Employing again the matching condition $\ell = |\mathbf{q}| / \mu$ to the RG 
equation solutions \eqref{eq:rg1} and \eqref{eq:rg2}, the universal scaling 
forms \eqref{eq:p1l} and \eqref{eq:c1l} become now generalized to
\begin{eqnarray} \label{eq:sca}
&&\chi^R(\mathbf{q},\omega,\{ p_R \},u_R)^{-1} = D_R^{-1}
\Gamma^R_{\tilde\psi \psi^*}(\mathbf{q},\omega,\{ p_R \},u_R) \nonumber \\
&&\qquad \approx \mu^\eta \, |\mathbf{q}|^{2 - \eta} 
\left( 1 + i r_{K R} \, |\mathbf{q} / \mu|^{\eta - \eta_c} \right) \\
&&\qquad\quad \times \, {\hat \Gamma}\left( \frac{\omega}{D_R \mu^{2 - z} \, 
|\mathbf{q}|^z \left( 1 + i r_{K R} \, |\mathbf{q} / \mu|^{\eta - \eta_c} \right)}
, \frac{\tau_R}{|\mathbf{q} / \mu|^{1 / \nu}} \right) \ , \nonumber \\
&&G^R_{\psi^* \psi}(\mathbf{q},\omega,\{ p_R \},u_R) \approx 
\frac{T_R}{D_R \mu^{2 - z + \eta'}} \, |\mathbf{q}|^{- 2 - z + \eta'} \nonumber \\ 
&&\qquad\quad \times \, {\hat C}\left( \frac{\omega}{D_R \mu^{2 - z} \, 
|\mathbf{q}|^z} , \frac{\tau_R}{|\mathbf{q} / \mu|^{1 / \nu}} , 
r_{K R} \, |\mathbf{q} / \mu|^{\eta - \eta_c} \right) \ ,
\end{eqnarray}
see Eqs.~\eqref{eq:chs} and \eqref{eq:cos}, that are valid in the vicinity of 
the critical point.
Here we have identified the set of universal scaling exponents as 
\begin{eqnarray} \label{eq:exp}
&&\nu^{-1} = - \gamma_\tau^* \ , \quad z = 2 + \gamma_D^* \ , \quad 
\eta = \zeta(u^*) - \gamma_D^* \ , \nonumber \\
&&\eta_c = \zeta(u^*) - \gamma_D^* - \gamma_{r_K}^* \ , \quad 
\eta' = 2 \Re \zeta(u^*) + \gamma_T^* \ . \quad
\end{eqnarray}
Note that $\eta - \eta_c = \gamma_{r_K}^*$ is determined by the anomalous 
scaling dimension of the non-equilibrium parameter $r_{K R}$.
Since $z - 2$, $\eta$, $\eta_c$, and $\eta'$ all vanish to first order in the
non-linear coupling, we only need to insert the stable one-loop equilibrium
fixed point value \eqref{eq:efp} to finally recover the standard equilibrium 
Model A critical exponents \eqref{eq:eta} and \eqref{eq:exz} to two-loop order:
\begin{eqnarray} \label{eq:2ex}
&&\eta = \frac{{u^*}^2}{18} + O({u^*}^3) = \frac{\epsilon^2}{50} 
+ O(\epsilon^3) \ , \nonumber \\
&&z = 2 + \frac{{u^*}^2}{18} \left( 6 \ln \tfrac43 - 1 \right) + O({u^*}^3) \\
&&\;\,\ = 2 + \frac{\epsilon^2}{50} \left( 6 \ln \tfrac43 - 1 \right) 
+ O(\epsilon^3) \ . \nonumber 
\end{eqnarray}
 Since the system has reached an {\em equilibrium} fixed point 
and is effectively thermalized, the identity \eqref{eq:eqi} immediately 
implies the {\em exact} scaling relation 
\begin{equation} \label{eq:etai}
\eta' = \zeta(u^*) - \gamma_D^* = \eta \ ,
\end{equation}
reflecting the emergence of detailed balance and the ensuing 
fluctuation-dissipation theorem.
This leaves us with a {\em single} new non-equilibrium drive scaling exponent 
\begin{eqnarray}
&&\eta_c = - \frac{{u^*}^2}{18} \left( 4 \ln \tfrac43 - 1 \right) + O({u^*}^3) 
\nonumber \\
&&\quad\!\ = - \frac{\epsilon^2}{50} \left( 4 \ln \tfrac43 - 1 \right) 
+ O(\epsilon^3) \ ,
\end{eqnarray}

If we (daringly) set $\epsilon = 1$, we find the critical exponents
$\nu \approx 0.625$ (to one-loop order), $\eta = \eta' \approx 0.02$, 
$z \approx 2 + 0.72609 \, \eta \approx 2.01452$, and
$\eta_c \approx - 0.15073 \, \eta \approx - 0.0030146$.
For comparison, the numerical values found in $d = 3$ dimensions by means of 
the non-perturbative RG approach in 
Refs.~\onlinecite{sieberer13:_dynam,sieberer13_long} are $\nu \approx 0.716$, 
$\eta \approx 0.039$, $z \approx 2.121$, and $\eta_c \approx - 0.223$.
The two-loop values thus  apparently still underestimate the 
fluctuation corrections in three dimensions.

\subsection{Observability of the drive exponent}

The fact that the drive exponent appears in the scaling form of the 
single-particle dynamical response makes it accessible to experimental 
observation \cite{sieberer13:_dynam,sieberer13_long}. 
In particular, the imaginary part of the dynamical response is probed in 
radiofrequency spectroscopy in ultracold atoms \cite{Stewart2008}, or homodyne 
detection for exciton-polariton systems \cite{utsu08} (in the latter, also the 
real part is available separately). 
For probe frequency $\omega$, the scaling form \eqref{eq:chs} implies for the response at criticality
\begin{equation} \label{eq:imr}
\chi(\mathbf{q},\omega) \approx \frac{1}{|\mathbf{q}|^{2 - \eta - z}}\, \frac{1}
{-i \left( \omega / D - a |\mathbf{q}|^{z + \eta - \eta_c} \right) + |\mathbf{q}|^z} \ ,
\end{equation}
which demonstrates different critical wavevector scaling for the peak position 
$\omega_0 \sim |\mathbf{q}|^{z + \eta - \eta_c}$ and peak width 
$\sim |\mathbf{q}|^z$.

\subsection{Complex Model B dynamics}

We conclude our field-theoretical approach with a brief RG analysis of a variant 
of the Langevin equation \eqref{eq:gpe3} with conserved order parameter 
dynamics, i.e., off-critical diffusive relaxation (Model B):
\begin{equation} \label{eq:ble}
\partial_t \psi(\mathbf{x},t) = D \nabla^2  
\frac{\delta \bar H[\psi]}{\delta \psi^*(\mathbf{x},t)} + \xi(\mathbf{x},t) \ ,
\end{equation}
with the non-Hermitean ``Hamiltonian'' \eqref{eq:ham}, and the noise 
correlations 
\begin{eqnarray} \label{eq:bns}
&&\left\langle \xi^*(\mathbf{x},t) \right\rangle 
= \left\langle \xi(\mathbf{x},t) \right\rangle = 0 \ , \nonumber \\
&&\left\langle \xi^*(\mathbf{x},t) \, \xi(\mathbf{x}',t') \right\rangle 
= - \gamma \nabla^2 \delta(\mathbf{x} - \mathbf{x}') \, \delta(t - t') \ , \\ 
&&\left\langle \xi^*(\mathbf{x},t) \, \xi^*(\mathbf{x}',t') \right\rangle 
= \left\langle \xi(\mathbf{x},t) \, \xi(\mathbf{x}',t') \right\rangle  = 0 
\ , \nonumber
\end{eqnarray}
 with $\gamma = 4 D T$
such that the standard $O(2)$-symmetric equilibrium Model B critical dynamics 
is incorporated as the special case with $r' = r_K = r_U = 0$.
When these parameters are non-zero, we have a conserved complex 
Ginzburg-Landau equation, or a driven, non-equilibrium version of the noisy 
Cahn-Hilliard equation with complex coefficients.

As a consequence of the conserved dynamics, both the Onsager relaxation 
coefficient in Eq.~\eqref{eq:ble} and the noise correlator strength 
\eqref{eq:bns} now carry an additional Laplacian operator.
This increases the mean-field dynamic critical exponent to $z = 4$, and 
generates an external wavevector factor $\mathbf{q}^2$ attached to the outgoing 
legs in the non-linear relaxation vertices depicted in Fig.~\ref{fg:fig1}(c).
Therefore, one has to {\em all} orders in the perturbation expansion
\begin{eqnarray}
&&\Gamma_{\tilde\psi \psi^*}(\mathbf{q}=0,\omega) = - i \omega \ , \nonumber \\
&&\partial_{q^2} \Gamma_{\tilde\psi \tilde\psi^*}(\mathbf{q},\omega=0) 
|_{{\mathbf q}=0} = - 2 D T \ .
\end{eqnarray}
Since these vertex functions carry no UV singularities, thus $Z = 1$ and 
 $Z_T = Z_D^{-1}$, which imply the {\em exact} results 
\begin{equation} \label{eq:bzg}
\zeta = 0 \ , \quad \gamma_T = - \gamma_D \ .
\end{equation}
Note that these Model B relations represent a special case of the
general equilibrium condition \eqref{eq:eqi}.
In addition, according to Eq.~\eqref{eq:zf1} $Z = 1$ enforces to two-loop 
order that $Z_{r_K} = 1 + O(u^3)$, whence $\gamma_{r_K} = 0 + O(u_R^3)$.

For the dynamical susceptibility $\chi(\mathbf{q},\omega) = D \mathbf{q}^2 
\Gamma_{\tilde\psi \psi^*}(\mathbf{q},\omega)^{-1}$ and correlation function
$C(\mathbf{q},\omega) = G_{\psi^* \psi}(\mathbf{q},\omega)$, one is led again 
to the scaling laws \eqref{eq:chs} and \eqref{eq:cos};
\begin{eqnarray} \label{eq:scb}
&&\chi^R(\mathbf{q},\omega,\{ p_R \},u_R)^{-1} = (D_R \mathbf{q}^2)^{-1}
\Gamma^R_{\tilde\psi \psi^*}(\mathbf{q},\omega,\{ p_R \},u_R) \nonumber \\
&&\qquad \approx \mu^\eta \, |\mathbf{q}|^{2 - \eta} 
\left( 1 + i r_{K R} \, |\mathbf{q} / \mu|^{\eta - \eta_c} \right) \\
&&\qquad\quad \times \, {\hat \Gamma}\left( \frac{\omega}{D_R \mu^{4 - z} \, 
|\mathbf{q}|^z \left( 1 + i r_{K R} \, |\mathbf{q} / \mu|^{\eta - \eta_c} 
\right)} , \frac{\tau_R}{|\mathbf{q} / \mu|^{1 / \nu}} \right) \ , \nonumber \\
&&G^R_{\psi^* \psi}(\mathbf{q},\omega,\{ p_R \},u_R) \approx 
\frac{T_R}{D_R \mu^{4 - z + \eta'}} \, |\mathbf{q}|^{- 2 - z + \eta'} 
\nonumber \\ 
&&\qquad\quad \times \, {\hat C}\left( \frac{\omega}{D_R \mu^{4 - z} \, 
|\mathbf{q}|^z} , \frac{\tau_R}{|\mathbf{q} / \mu|^{1 / \nu}} , 
r_{K R} \, |\mathbf{q} / \mu|^{\eta - \eta_c} \right) \ .
\end{eqnarray}
Here, the static critical exponents $\nu = - 1 / \gamma_\tau^*$ and 
$\eta = - \gamma_D^*$ are identical to the values \eqref{eq:1nu} and 
\eqref{eq:eta} computed in Secs.~\ref{sec:renor}.C and E. 
In contrast to its Model A counterpart, however, we now have within the 
two-loop approximation
\begin{equation} \label{eq:bed}
\eta_c = \eta + O(\epsilon^3) = \frac{\epsilon^2}{50} + O(\epsilon^3) \ .
\end{equation}
Yet the relations \eqref{eq:bzg} imply for the complex non-equilibrium Model B 
that to {\em all} orders in perturbation theory
\begin{equation} \label{eq:bex}
z = 4 + \gamma_D^* = 4 - \eta \ , \quad \eta' 
= \gamma_T^* = \eta \ .
\end{equation}
The dynamic critical exponent thus assumes its standard equilibrium value, as 
for the driven-dissipative extension of Model A.
In addition, $\eta' = \eta$ is confirmed exactly here, as opposed to the 
situation in Model A, and to Eq.~\eqref{eq:bed}, for which we can establish 
this identity only up to $O(\epsilon^3)$ corrections.
This shows that unlike the driven-dissipative extension of Model A, no 
independent critical behavior is found for an analogous extension of Model 
B, at least to two-loop order.

\section{Conclusion and Outlook}
\label{sec:concl}

We have obtained a detailed picture of the driven-dissipative Bose condensation 
transition using the well-developed framework of the perturbative 
field-theoretical dynamic renormalization group, in this way complementing a 
previous functional renormalization group study 
\cite{sieberer13:_dynam,sieberer13_long}. 
In particular, we traced back the existence of a new, independent critical 
exponent to additional UV divergences without counterpart in the critical theory
for the equilibrium Bose condensation transition, and obtained its analytical 
value at two-loop order in the dimensional $\epsilon$ expansion. 
This exponent is present in the scaling form of the dynamic single-particle 
response of the system,  and in its two-point correlation 
functions, and witnesses non-equilibrium conditions at the largest distances 
in the problem. 
We furthermore confirmed explicitly the asymptotic thermalization scenario in 
the renormalization group flow for this system, as in Ref.~\onlinecite{risler05} 
demonstrating the stability of the equilibrium fixed point, along with the 
construction of the corresponding dynamical scaling forms and calculation of the
associated critical exponents.

The perturbative field theoretic approach offers the possibility for a direct 
comparison of the results for non-equilibrium systems with well-studied 
counterparts in thermodynamic equilibrium. 
This makes it a valuable tool for future investigations of critical 
driven-dissipative  quantum systems, working towards the ultimate 
goal of a systematic classification of non-equilibrium dynamic criticality to 
a similar maturity level as has been achieved in thermodynamic equilibrium. 
In the present context, first steps include the exploration of 
different internal symmetries, such as general $O(n)$ rotation invariance, or 
the inclusion of explicit coherent pumping processes.  
They also comprise an investigation of universal aspects of the dynamics 
following a parameter quench, such as the determination of the initial slip 
exponent and critical aging \cite{janssen89,calabrese05} in driven-dissipative 
systems. 
It remains to be seen whether this approach can also be leveraged to situations 
where criticality and genuine quantum effects come into play simultaneously.

\appendix

\section{Integrals and technical details}
\label{sec:appen}

\subsection{Dimensionally regularized integrals}

In this Appendix, we provide a list of wavevector integrals, of 
the form \eqref{eq:wvi}, evaluated in dimensional regularization 
\cite{amitbook,itzyksonbook,zinnjustinbook,kleinert-book,tauberbook}.
The UV divergences at $d_c = 4$ become manifest as simple poles in 
$\epsilon = 4 - d$.
The basic dimensionally regularized integral is
\begin{equation} \label{eq:dmr}
\int \! \frac{d^dk}{(2 \pi)^d} \, 
\frac{1}{\left( \tau + 2 \, \mathbf{q} \cdot \mathbf{k} + \mathbf{k}^2 
\right)^s} = \frac{\Gamma(s - d/2)}{(4 \pi)^{d/2} \Gamma(s)} \, 
\frac{1}{\left( \tau - q^2 \right)^{s - d/2}} \ .
\end{equation}
This immediately gives the wavevector integrals required for the one-loop 
analysis, with $A_d = \Gamma(3 - d/2) / 2^{d-1} \pi^{d/2}$:
\begin{eqnarray}
&&\int_k \frac{1}{\tau + k^2} = 
- \, \frac{2 A_d}{(d - 2) \, (4 - d)} \, \tau^{-1 + d/2} \ , \label{eq:dr1} \\
&&\int_k \frac{1}{\left( \tau + k^2 \right) \left( \tau' + k^2 \right)} = 
\frac{2 A_d}{(d - 2) \, (4 - d)} \, 
\frac{\tau^{-1 + d/2} - {\tau'}^{-1 + d/2}}{\tau - \tau'} \ , \nonumber \\
&&\label{eq:dr2} \\
&&\int_k \frac{1}{\left( \tau + k^2 \right)^2} = 
\frac{A_d}{4 - d} \, \tau^{-2 + d/2} \ . \label{eq:dr3}
\end{eqnarray}

In order to evaluate the nested wavevector integrals appearing in the two-loop 
calculation, Feynman's parametrization is very useful:
\begin{equation} \label{eq:fep}
\frac{1}{A^r B^s} = \frac{\Gamma(r + s)}{\Gamma(r) \, \Gamma(s)} 
\int_0^1 \! \frac{x^{r-1} \, (1 - x)^{s-1}}{[x A + (1 - x) B]^{r+s}} \, dx \ . 
\end{equation} 
We first extract the UV-singular part of the two-loop integral \eqref{eq:dqt}, 
a standard textbook computation \cite{tauberbook}. 
Feynman's parametrization and the $\mathbf{p}$ integration by means of 
Eq.~\eqref{eq:dmr} yield 
\begin{eqnarray} \label{eq:dq1}
&&D_d\left( \tau, \mathbf{q} \right) = \int_0^1 \! dx 
\int_k \frac{1}{\tau + k^2} \nonumber \\
&&\qquad\quad \times \int_p \frac{1}{\left[ \tau + x \left( 
\mathbf{q} - \mathbf{k} \right)^2 - 2 x \left( \mathbf{q} - \mathbf{k} \right) 
\cdot \mathbf{p} + \mathbf{p}^2 \right]^2} \\ 
&&= \frac{\Gamma(2 - d/2)}{(4 \pi)^{d/2}} \! \int_0^1 \!\! dx 
\int_k \frac{1}{\tau + k^2} \, \frac{1}{\left[ \tau + x (1 - x) 
\left( \mathbf{q} - \mathbf{k} \right)^2 \right]^{2 - d/2}} \, . \nonumber
\end{eqnarray} 
Applying \eqref{eq:fep} once more, the $\mathbf{k}$ integral can be performed, 
\begin{eqnarray} \label{eq:dq2} 
&&D_d\left( \tau, \mathbf{q} \right) = \frac{\Gamma(3 - d/2)}{(4 \pi)^{d/2}} 
\int_0^1 \! \frac{dx}{[x (1 - x)]^{2 - d/2}} \nonumber \\ 
&&\ \times \int_0^1 \! dy \int_k \frac{y^{1 - d/2}}{\left[ \tau \left( 
\frac{y}{x (1 - x)} + 1 - y \right) + y \, \mathbf{q}^2 - 2 y \, \mathbf{q} 
\cdot \mathbf{k} + k^2 \right]^{3 - d/2}} \nonumber \\ 
&&\quad = \frac{\Gamma(3 - d)}{(4 \pi)^d} \int_0^1 \! 
\frac{dx}{[x (1 - x)]^{2 - d/2}} \nonumber \\
&&\qquad\quad \times \int_0^1 \! \frac{y^{1 - d/2} \, dy}{\left[ \tau \, 
\Bigl( \frac{y}{x (1 - x)} + 1 - y \Bigr) + y (1 - y) \, q^2 \right]^{3 - d}} 
\, . 
\end{eqnarray} 
Therefore one obtains in $d = 4 - \epsilon$ dimensions
\begin{eqnarray} \label{eq:dq3} 
&&\partial_{q^2} D_d\left( \tau, \mathbf{q} \right) |_{\mathbf{q}=0} = 
- \frac{\Gamma(1 + \epsilon)}{\Gamma(1 + \epsilon / 2)^2} \, 
\frac{A_d^2 \, \tau^{- \epsilon}}{4 \, \epsilon} \!
\int_0^1 \! \frac{dx}{[x (1 - x)]^{\epsilon / 2}} \nonumber \\ 
&&\qquad\quad \times \int_0^1 \! dy \, y^{- \epsilon / 2} \, (1 - y) \,
\biggl[ \frac{y}{x (1 - x)} + 1 - y \biggr]^{- \epsilon} \, .
\end{eqnarray} 
Noting that $\Gamma(1 + \epsilon) / \Gamma(1 + \epsilon / 2)^2 
= 1 + O(\epsilon^2)$, and that the parameter integrals are regular in the limit 
$\epsilon \to 0$, one finally isolates the $1 / \epsilon$ pole and its residuum
\begin{equation} \label{eq:dq4} 
\partial_{q^2} D_d\left( \mu^2,\mathbf{q} \right) 
\big|^{\rm sing.}_{\mathbf{q}=0} = -
\frac{A_d^2 \, \mu^{- 2 \epsilon}}{4 \, \epsilon} \! \int_0^1 \! (1 - y) \, dy
= - \frac{A_d^2 \, \mu^{- 2 \epsilon}}{8 \, \epsilon} \ . 
\end{equation}

\begin{widetext}
For the integral \eqref{eq:idt}, we proceed similarly:
Feynman's parametrization leads to
\begin{eqnarray} \label{eq:id1}
&&I_d\left( \tau, r_K, r_U \right) = \! \int_0^1 \! dx \! \int_0^1 \! dy \, y 
\! \int_k \frac{1}{\tau + k^2} \int_p \frac{1}{\left( \tau \left[ 1 + \tfrac12 
(1 - y) (1 + i r_U) \right] + \left( x y + 1 - y \right) k^2 + 2 \left[ x y 
+ \tfrac12 (1 - y) (1 - i r_K) \right] \mathbf{k} \cdot \mathbf{p} + p^2 
\right)^3} \nonumber \\
&&\qquad 
= \frac{\Gamma(3 - d/2)}{2 (4 \pi)^{d/2}} \int_0^1 \! dx \int_0^1 \! dy \, y \\
&&\qquad\quad \times \int_k \frac{1}{\tau + k^2} \, \frac{1}{\left( \tau 
\left[ 1 + \tfrac12 (1 - y) (1 + i r_U) \right] + \left( x y \left[ 1 - x y 
- (1 - y) (1 - i r_K) \right] + (1 - y) \left[ 1 - \tfrac14 (1 - y) 
(1 - i r_K)^2 \right] \right) k^2 \right)^{3 - d/2}} \ , \nonumber 
\end{eqnarray}
after performing the $\mathbf{p}$ integral by means of  Eq.~\eqref{eq:dmr}.
Employing Eq.~\eqref{eq:fep} once more and subsequently carrying out the 
$\mathbf{k}$ integral gives
\begin{eqnarray} \label{eq:id2}
&&I_d\left( \tau, r_K, r_U \right) = 
\frac{\Gamma(4 - d) \, \tau^{d-4}}{2 (4 \pi)^{d/2}} \int_0^1 \! dx \int_0^1 \! 
dy \, \frac{y}{\left[ x (1 - x) y^2 + x y (1 - y) i r_K + \tfrac14 (1 - y) 
\left[ 3 + y + (1 - y) r_K^2 \right] + 2 (1 - y) i r_K \right]^{3 - d/2}} 
\nonumber \\ 
&&\ \times \int_0^1 \! dz \, \frac{z^{2 - d/2}}{\left( 1 - z + z \left[ 3 - y 
+ (1 - y) i r_U \right] / \left[ 2 x (1 - x) y^2 + \tfrac12 (1 - y) \left[ 
3 + y + (1 - y) r_K^2 \right] + 2 (1 - y) \left[ x y + \tfrac12 (1 - y) i r_K
\right] i r_K \right] \right)^{4 - d}} \ .\nonumber\\
\end{eqnarray}
Evaluated in $d = 4 - \epsilon$ dimensions, the $z$ parameter integral is just
$1 + O(\epsilon^2)$, whence  Eq.~\eqref{eq:id2} reduces to
\begin{equation} \label{eq:id3}
I_d\left( \mu^2,r_K \right) \big|^{\rm sing.} = 
\frac{A_d^2 \, \mu^{- 2 \epsilon}}{2 \epsilon} \int_0^1 \! dx \int_0^1 \! dy \,
\frac{y}{4 x (1 - x) y^2 + (1 - y) \left[ 3 + y + (1 - y) r_K^2 \right] 
+ 2 (1 - y) (2 x y + 1 - y) \, i r_K} \ , 
\end{equation}
which remarkably comes out independent of the parameter $r_U$.

\subsection{Complex parameter integral}

The integration over $x$ in  Eq.~\eqref{eq:id3} is elementary, and gives
\begin{eqnarray} \label{eq:id4}
&&I_d\left( \mu^2, r_K \right) \big|^{\rm sing.} = 
\frac{A_d^2 \, \mu^{- 2 \epsilon}}{8 \epsilon} \int_0^1 \! dy \, 
\frac{1}{\sqrt{3 - 2 y + 2 (1 - y) \, i r_K}} \nonumber \\
&&\qquad\qquad \times \ln \, \Biggl( 
\frac{- y + (1 - y) \, i r_K - \sqrt{3 - 2 y + 2 (1 - y) \, i r_K}}
{- y + (1 - y) \, i r_K + \sqrt{3 - 2 y + 2 (1 - y) \, i r_K}} \, 
\frac{y + (1 - y) \, i r_K + \sqrt{3 - 2 y + 2 (1 - y) \, i r_K}}
{y + (1 - y) \, i r_K - \sqrt{3 - 2 y + 2 (1 - y) \, i r_K}} \, \Biggr) \ .
\end{eqnarray}
After substitution to the new integration variable 
$z^2 = 3 - 2 y + 2 (1 - y) \, {\tilde r}$, where ${\tilde r} = i r_K$ is
treated as an arbitrary parameter, the integral becomes
\begin{equation} \label{eq:id5}
I_d\left( \mu^2, r_K \right) \big|^{\rm sing.} = 
\frac{A_d^2 \, \mu^{- 2 \epsilon}}{8 \epsilon \, (1 + {\tilde r})} 
\int_1^{\sqrt{3 + 2 {\tilde r}}} \! dz \, \Biggl[ 2 \ln \frac{z + 1}{z - 1} 
+ \ln \frac{3 - z}{3 + z} + \ln \frac{3 + {\tilde r} - (1 - {\tilde r}) \, z}
{3 + {\tilde r} + (1 - {\tilde r}) \, z} \Biggr] \ .
\end{equation}
Evaluating these again elementary integrals, and returning to $i r_K$ through
analytic continuation into the complex plane, one at last arrives at
\begin{equation} \label{eq:id6}
I_d\left( \mu^2, r_K \right) \big|^{\rm sing.} = 
\frac{A_d^2 \, \mu^{- 2 \epsilon}}{8 \epsilon \, (1 + i r_K)} \, \Biggl[ 2 \, 
\frac{3 - i r_K}{1 - i r_K} \ln \frac{4}{3 - i r_K} - 2 \, 
\frac{1 + i r_K}{1 - i r_K} \ln (1 + i r_K) + \sqrt{3 + 2 i r_K} \, L(r_K)
\Biggr] \ ,
\end{equation}
with the logarithm of three products of complex ratios
\begin{eqnarray} \label{eq:lrk}
&&L(r_K) = \ln \, \Biggl( \frac{\sqrt{3 + 2 i r_K} + 1}{\sqrt{3 + 2 i r_K} - 1} 
\, \frac{3 - \sqrt{3 + 2 i r_K}}{3 + \sqrt{3 + 2 i r_K}} \, 
\frac{\sqrt{3 + 2 i r_K} + i r_K}{\sqrt{3 + 2 i r_K} - i r_K} \Biggr) \nonumber
\\
&&\qquad\;\ = \tfrac12 \ln \, \Biggl( 
\frac{1 + 3 s_K + \sqrt{6} \sqrt{s_K + 1}}{1 + 3 s_K - \sqrt{6} \sqrt{s_K + 1}} 
\, \frac{3 + s_K - \sqrt{6} \sqrt{s_K + 1}}{3 + s_K + \sqrt{6} \sqrt{s_K + 1}} 
\, \frac{3 s_K + r_K^2 + \sqrt{6} \, r_K \sqrt{s_K - 1}}
{3 s_K + r_K^2 - \sqrt{6} \, r_K \sqrt{s_K - 1}} \, \Biggr) \ ,
\end{eqnarray}
where $s_k = \sqrt{1 + \tfrac49 \, r_K^2}$.
Note that the complex phases remarkably cancel in  Eq.~\eqref{eq:lrk}, leaving
a real expression; furthermore, $L(0) = 0$.
We specifically require
\begin{equation} \label{eq:ir1}
\left( 1 + r_K^2 \right) {\rm Re} \, I_d\left( \mu^2, r_K \right)
\big|^{\rm sing.}\!\! = \frac{A_d^2 \, \mu^{- 2 \epsilon}}{8 \epsilon} \, 
\Biggl[ 3 \ln \frac{16}{9 + r_K^2} - \ln \left(1 + r_K^2 \right) 
+ 2 r_K \left( \arctan r_K + \arctan \frac{r_K}{3} \right) + \sqrt{\tfrac32} 
\left( \! \sqrt{s_K + 1} + r_K \sqrt{s_K - 1} \right) L(r_K) \Biggr] \, ,
\end{equation}
and
\begin{eqnarray} \label{eq:ir2}
&&\!\!\!\!\!\! \left( 1 + i r_K \right) I_d\left( \mu^2, r_K \right) 
\big|^{\rm sing.} \!\! = \frac{A_d^2 \, \mu^{- 2 \epsilon}}{8 \epsilon} \, 
\Biggl[ \frac{3 + r_K^2}{1 + r_K^2} \ln \frac{16}{9 + r_K^2} 
- \frac{1 - r_K^2}{1 + r_K^2} \ln \left( 1 + r_K^2 \right) 
+ \frac{4 r_K}{1 + r_K^2} \left( \arctan r_K - \arctan \frac{r_K}{3} \right) 
+ \sqrt{\tfrac32 \left( s_k + 1 \right)} \, L(r_K) \Biggr] \nonumber \\
&&\qquad\qquad\quad + i \, \frac{A_d^2 \, \mu^{- 2 \epsilon}}{8 \epsilon} \, 
\Biggl[ \frac{2 r_K}{1 + r_K^2} \, \ln \frac{16}{(9 + r_K^2) \, (1 + r_K^2)} 
- 2 \, \frac{1 - r_K^2}{1 + r_K^2} \arctan r_K 
+ 2 \, \frac{3 + r_K^2}{1 + r_K^2} \arctan \frac{r_K}{3} 
+ \sqrt{\tfrac32 \left( s_K - 1 \right)} \, L(r_K) \Biggr] \ .
\end{eqnarray}
\end{widetext}

\acknowledgments
We thank Ehud Altman, Michel Pleimling, Lukas Sieberer, and Royce Zia for 
insightful discussions, and Hiba Assi for graphing Fig.~\ref{fg:fig6}.
This research was in part (S.D.) supported through the Austrian Science Fund 
(FWF) with START Grant No. Y 581-N16 and SFB FoQuS (FWF Project No. F4006-N16).

\bibliography{bibliography}
  
\end{document}